\documentstyle[12pt,epsf]{article}  
\def\thefootnote{\fnsymbol{footnote}}
\def\beq{\begin{equation}}
\def\eeq{\end{equation}}
\def\to{\rightarrow}

\def\bsg{\ifmmode B\to X_s\gamma\else $B\to X_s\gamma$\fi}
\def\bsll{\ifmmode B\to X_s\ell^+\ell^-\else $B\to X_s\ell^+\ell^-$\fi}
\def\bstt{\ifmmode B\to X_s\tau^+\tau^-\else $B\to X_s\tau^+\tau^-$\fi}
\def\shat{\ifmmode \hat{s}\else $\hat{s}$\fi}

\newcommand{\newc}{\newcommand}

\newc{\lcal}{\int {\cal L}dt}
 
\newc{\LSP}{{\chi^0_1}}
\newc{\stauR}{{\tilde \tau_R}}
\newc{\stau}{{\tilde \tau_1}}
\newc{\mstop}{m_{\tilde{t}}}
\newc{\mHpm}{m_{H^\pm}}
\newc{\gsim}{\lower.7ex\hbox{$\;\stackrel{\textstyle>}{\sim}\;$}}
\newc{\lsim}{\lower.7ex\hbox{$\;\stackrel{\textstyle<}{\sim}\;$}}
\newc{\ie}{{\it i.e.}}          
\newc{\etal}{{\it et al.}}
\newc{\eg}{{\it e.g.}}          
\newc{\kev}{\hbox{\rm\,keV}}            
\newc{\mev}{\hbox{\rm\,MeV}}            
\newc{\gev}{\hbox{\rm\,GeV}}            
\newc{\tev}{\hbox{\rm\,TeV}}
\newc{\xpb}{\hbox{\rm\, pb}}
\newc{\xfb}{\hbox{\rm\, fb}}

\newc{\mtop}{m_t}
\newc{\mbot}{m_b}
\newc{\mz}{m_Z}
\newc{\mw}{M_W}
\newc{\alphasmz}{\alpha_s(m_Z^2)}
\newc{\swsq}{\sin^2\theta_W}
\newc{\tw}{\tan\theta_W}
\newc{\cw}{\cos\theta_W}
\newc{\sw}{\sin\theta_W}
\newc{\BR}{\hbox{\rm BR}}
\newc{\zbb}{Z\to b\bar}
\newc{\Gb}{\Gamma (Z\to b\bar b)}
\newc{\Gh}{\Gamma (Z\to \hbox{\rm hadrons})}
\newc{\rbsm}{R_b^\hbox{\rm sm}}
\newc{\rbsusy}{R_b^\hbox{\rm susy}}
\newc{\drb}{\delta R_b}

\newc{\sgn}{\mbox{sgn}}

\def\md{M_D}
\def\sch{Schwarzschild }
\def\eq#1{eq.~(\ref{#1})}
\def\beqa{\begin{eqnarray}}
\def\eeqa{\end{eqnarray}}
\def\slash#1{\not  \! \! {#1}}

\def\beq{\begin{equation}}
\def\eeq{\end{equation}}
\def\bea{\begin{eqnarray}}
\def\eea{\end{eqnarray}}
\def\slashchar#1{\setbox0=\hbox{$#1$}           
   \dimen0=\wd0                                 
   \setbox1=\hbox{/} \dimen1=\wd1               
   \ifdim\dimen0>\dimen1                        
      \rlap{\hbox to \dimen0{\hfil/\hfil}}      
      #1                                        
   \else                                        
      \rlap{\hbox to \dimen1{\hfil$#1$\hfil}}   
      /                                         
   \fi}                                         %
\catcode`@=11
\long\def\@caption#1[#2]#3{\par\addcontentsline{\csname
  ext@#1\endcsname}{#1}{\protect\numberline{\csname
  the#1\endcsname}{\ignorespaces #2}}\begingroup
    \small
    \@parboxrestore
    \@makecaption{\csname fnum@#1\endcsname}{\ignorespaces #3}\par
  \endgroup}
\catcode`@=12

\def\dofig#1#2{\epsfxsize=#1\centerline{\epsfbox{#2}}}
\def\dofigs#1#2#3{\centerline{\epsfxsize=#1\epsfbox{#2}%
   \hfil\epsfxsize=#1\epsfbox{#3}}}
\def\jfig#1#2#3{
 \begin{figure}
\begin{centering}
 \epsfysize=2.8in
 \centerline{\epsfbox{#2}}
 \caption{#3}
 \label{#1}
\end{centering}
 \end{figure}}
\def\sfig#1#2#3{
 \begin{figure}
 \centering
 \epsfysize=2.1in
 \hspace*{0in}
 \centerline{\epsfbox{#2}}
 \caption{#3}
 \label{#1}
 \end{figure}}

\begin{document}

\baselineskip=18pt

\begin{titlepage}
\begin{flushright}
CERN-TH/2001-306\\
DESY 01-206 \\
hep-ph/0112161
\end{flushright}

\begin{center}
\vspace{1cm}

{\Large \bf 

Transplanckian Collisions at the LHC and Beyond}

\vspace{0.8cm}

{\bf Gian F. Giudice$^a$, Riccardo Rattazzi$^a$\footnote{On leave of absence
from INFN, Pisa.}, James D. Wells$^{b,c}$}

\vspace{.5cm}

{\it ${}^{(a)}$CERN, Theory Division, CH-1211 Geneva 23, Switzerland}\\
{\it ${}^{(b)}$Physics Department, University of California, Davis, CA 95616,
USA}\\
{\it ${}^{(c)}$Deutsches Elektronen-Synchrotron DESY, D-22603 Hamburg, Germany}

\end{center}
\vspace{1cm}

\begin{abstract}
\medskip
Elastic collisions in the transplanckian region, where the center-of-mass 
energy is much larger than the fundamental gravity mass scale, can be 
described by linearized general relativity and known quantum-mechanical 
effects as long as the momentum transfer of the process is sufficiently 
small. For larger momentum transfer, non-linear gravitational effects 
become important and, although a computation is lacking, black-hole 
formation is expected to dominate the dynamics. We discuss how elastic 
transplanckian collisions can be used at high-energy colliders to study, 
in a quantitative and model-independent way, theories in which gravity 
propagates in flat extra dimensions. At LHC energies, however, 
incalculable quantum-gravity contributions may significantly affect the 
experimental signal.
 
\end{abstract}

\bigskip
\bigskip

\begin{flushleft}
December 2001
\end{flushleft}

\end{titlepage}

\tableofcontents
\vfill\eject

\addtocounter{footnote}{-1}
\def\thefootnote{\arabic{footnote}}
\setcounter{footnote}{0}
\section{Introduction}
\label{sec1}

The hypothesis that the characteristic energy scale of quantum gravity
lies just beyond the electroweak scale finds its motivation in the hierarchy
problem and its operative realization in the existence of extra spatial
dimensions~\cite{add,rs}. If this hypothesis is verified in nature, future
experiments at the LHC have the ground-breaking opportunity of testing the 
dynamics of
gravitation in the high-energy and quantum regimes. Unfortunately, since
quantum-gravity dynamics
are still unknown, systematic theoretical predictions cannot be made.
It is then important to identify certain collider
topologies and kinematical regions that allow a description based 
only upon general principles and
not on model peculiarities, 
and which can
be used in the experimental program as a handle to demonstrate
the gravitational nature
of the interaction. Single graviton emissions satisfy these requirements and
allow for a model-independent study of the propagation of gravitational forces
in extra spatial dimensions~\cite{noi,altri}. 
On the other hand, model-dependent experimental signals, as
those originating from contact interactions~\cite{noi,kkgra}
or, more likely,  the observation of a specific spectrum of
string Regge
excitations~\cite{pesk} will be very important
to discriminate among various models and to guide theoretical research. 

The model-independent experimental signals discussed in refs.~\cite{noi,altri}
are based upon interactions at 
energies below the quantum gravity scale $\md$ (the analogue of the Planck
mass for a $D$-dimensional theory), and their theoretical description
fails as we approach $\md$. However, there is another kinematical
regime that can be tackled in a fairly model-independent fashion: the
transplanckian region $\sqrt{s}\gg \md$. This is the subject of this 
paper. A pecularity of hadron machines is that, because of the composite
structure of the proton, one can study, in the same environment, parton 
collisions at different center-of-mass energies. Consequently, 
if $\md\simeq\tev$ experiments
at the LHC could probe the cisplanckian (graviton emission,
contact interactions), the planckian (the core of the quantum-gravity
dynamics), and the transplanckian regions.

To understand the nature of transplanckian collisions, we first identify 
the relevant scales (see ref.~\cite{venezia}). 
Let us consider general relativity in $D=4+n$
dimensions
with (generalized) Newton's constant $G_D$. For $n$ 
flat dimensions
compactified in a volume $V_n$, $G_D$ is related to the usual Newton's
constant by $G_D=V_n G_N$. 
To better elucidate the relation between the transplanckian and the
classical limit, we reinstate for a moment the correct powers of 
$\hbar$ and $c$. 
The relation between $G_D$ and the $D$-dimensional
Planck scale is 
\beq
G_D=\frac{(2\pi)^{n-1}\hbar^{n+1}}{4c^{n-1}M_D^{n+2}},
\eeq 
where the proportionality
constant follows from the conventions of ref.~\cite{noi}. Notice that
$G_D$ has dimensions $[G_D]=\ell^{n+5}E^{-1}t^{-4}$ ($\ell =$ length,
$E=$ energy, $t=$ time). 
Starting from $G_D$, we can construct the Planck length 
\beq
\lambda_P = \left( \frac{G_D \hbar}{c^3}\right)^{\frac{1}{n+2}}.
\label{lamp}
\eeq
This is the distance below which quantum gravity effects become important.

Using the center-of-mass energy of the collision, we can construct the
length
\beq
\lambda_B =\frac{4\pi \hbar c}{\sqrt{s}}.
\eeq
This is the de Broglie wavelength of the colliding particles, which  
characterizes their quantum length scale.

Combining $G_D$ and $\sqrt{s}$, we can form the \sch radius of a
system with center-of-mass energy $\sqrt{s}$~\cite{perry}
\beq
R_S =\frac{1}{\sqrt{\pi}}
\left[ \frac{8\Gamma \left( \frac{n+3}{2}\right)}{(n+2)}\right]^{
\frac{1}{n+1}}~ \left( \frac{G_D \sqrt{s}}{c^4}\right)^{\frac{1}{n+1}}.
\eeq
This is the length at which curvature effects become significant.
In the limit $\hbar \to 0$, with $G_D$ and $\sqrt s$ fixed,
$M_D$ vanishes,
showing that classical physics correspond to transplanckian 
(macroscopically large) energies. Moreover, in the same limit,
$R_S$ remains finite, while
the two length scales $\lambda_P$ and $\lambda_B$ go to zero.
Therefore, the transplanckian regime corresponds to a classical 
limit in which
the length scale $R_S$ characterizes the dynamics,
\beq
\sqrt{s} \gg M_D  ~~~~~~~\Rightarrow~~~~~~R_S\gg \lambda_P \gg \lambda_B.
\eeq
For instance, by simple dimensional analysis and analiticity in $G_D$,
we expect that the scattering angle for a collision with impact parameter 
$b$ is
$\theta\sim G_D\sqrt s/b^{n+1}=(R_S/b)^{n+1}$. This behavior shows
that by increasing the energy we can obtain a finite
angle scattering by going to larger $b$, and therefore further suppressing
short distance quantum gravity effects.

This property of gravity should be contrasted to what happens in an ordinary
gauge theory, where spin-1 particles mediate the force. In $(4+n)$--dimensional
space-time the gauge coupling $e^2$ has dimension $[e^2]=\ell^{n+1} E$.
The analogues of $M_D$, $\lambda_P$ and $R_S$ are given respectively by
$M_e^{n}=(\hbar c)^{n+1}/e^2$, $\lambda_e^n=e^2/(\hbar c)$ and 
$R_e^{n+1}=e^2/{\sqrt s}$. In the $\hbar \to 0$ limit, with
$e^2$ and $\sqrt s$ fixed, we find that $\lambda_e\to \infty$ (we focus
for simplicity on $n>0$)
indicating that quantum fluctutations are not suppressed
at any finite length. Indeed the same dimensional
argument we used before gives a classical scattering angle $\theta\sim
e^2/(\sqrt s b^{n+1})$, which remains finite as $\sqrt s\to \infty$ provided
$b\to 0$. But $b<\lambda_e$ is the regime where quantum effects 
are unsuppressed, so we conclude that there is no classical limit.
The different properties of spin-2 and spin-1 exchange that are elucidated
by this simple dimensional analysis are the same that render
the eikonal approximation consistent for the first and inconsistent
for the second. Basically it is because energy itself plays the role
of charge in gravity: when the energy is large, gravity
gives sizeable effects also at large distance where quantum
fluctuations of the geometry are irrelevant.

Having established that scattering at transplanckian energies is described by
classical physics, one also realizes that the corresponding amplitude 
should only be calculable by a non-perturbative approach. This is always
the case for semiclassical amplitudes (like instantons for example), for
which the classical action $S$ is much larger than $\hbar$, and 
we cannot perturbatively expand $\exp (-S/\hbar)$. 
In order to tackle the non-perturbative problem, we will further restrict
the range of distance scales under investigation, and only consider collisions
with impact parameter $b$ much larger than $R_S$. In this regime, the
curvature is small, the metric is nearly flat, and we can work 
in the limit of weak gravitational field,
neglecting non-linear effects of the graviton couplings. The 
non-perturbative interactions between the high-energy colliding partons and
the weak gravitational field can be computed using the eikonal
approximation, which can be trusted at small scattering angle.
From a field-theoretical point of view, this approximation 
corresponds~\cite{eik} to
a resummation of an infinite set of Feynman diagrams which, at each order
in perturbation theory, give the leading contributions to the forward
scattering. The eikonal approximation has been employed for 4-dimensional 
gravity~\cite{kab1}
and for string theory~\cite{venezia,sold} (see also ref.~\cite{areva}). 
The results for the scattering amplitude
agree with those obtained~\cite{hoof} by
solving the Klein-Gordon equation for a particle propagating in the
classical gravitational shock wave produced by the other 
particle~\cite{Aichelburg},
or by reformulating quantum gravity as a topological field theory~\cite{ver}.
All these different approaches give equivalent results~\cite{kab1,equiv}.

In sect.~\ref{sec2} we will derive and discuss the scattering amplitude
in the eikonal approximation for 4-dimensional colliding particles (living
on a 3-brane) and $D$-dimensional gravity~\cite{ratta} (see also 
ref.~\cite{nussi}). 
As explained above, the process is essentially classical, like the motion
of two (relativistic) stars coming within a close distance. 
However, an important 
difference is that the gravitational potential in the non-relativistic
analogue of our process decreases with the distance $r$ as $V(r)\propto
G_D/ r^{n+1}$. As discussed in sect.~\ref{sec2} this will have significant
consequences, and it will lead to quite different results from the case
of 4-dimensional Newtonian potentials.
A novel feature is the emergence of a new length scale
\beq
b_c \propto \left( \frac{G_D s}{\hbar c^5}\right)^{\frac{1}{n}}.
\eeq
Notice that $b_c$ cannot be defined in 4 dimensions ($n=0$). In the 
limit $\hbar \to 0$, we find that $b_c$ goes to infinity and therefore
the classical region extends only up to length scales of order $b_c$.
For impact
parameters larger than $b_c$, the process is no longer determined by
classical properties. However, since quantum gravity effects
are always negligible, ordinary relativistic quantum mechanics is
sufficient to compute scattering amplitudes within our approximation.

As previously discussed, we are limited by our computational ability to
work at $b \gg R_S$. As the impact parameter approaches the \sch
radius (and our expressions become unreliable), we expect to enter a
(classical) regime in which the gravitational field is strong, non-linear
effects are important and, plausibly, black holes are formed. 
This regime, discussed
in refs.~\cite{banks} and~\cite{blackh} (see also ref.~\cite{orel}), 
is very exciting from the experimental point of view,
but more arduous from a theoretical point of view. While in the eikonal
region we are able to make quantitatively reliable predictions with
controllable expansion parameters, the black-hole production cross section 
can only be estimated with dimensional arguments and more
precise information should wait for numerical simulations in $D$-dimensional
general relativity.

Black-hole production results in a black disk of radius $R_S$
in the cross section, since any initial state with $b<R_S$ is
absorbed by the black hole (this point of view has been criticized
in ref.~\cite{voloshin}). Collisions with
an impact parameter $b$ smaller than the \sch radius do not directly
probe physics at the 
energy scale $1/b$. Ultraviolet dynamics and quantum effects
are screened by the black hole.

The physical picture we have been describing holds if there is a clear
separation of the various length scales and if no other new dynamics   
intervene at distances between $R_S$ and $b_c$.
As we will discuss in this paper, this is not necessarily the case for 
particle collisions at the LHC. Given the present lower limits on $M_D$,
the center-of-mass energy at the LHC can be only marginally in
the transplanckian region, and a clear separation between the quantum-gravity
scale $\lambda_P$ and the classical scale $R_S$ is not completely achieved,
endangering our approximations. The problem is particularly apparent in
string theory, which contains a new scale, the string length
\beq
\lambda_S = \sqrt{\alpha^\prime},
\eeq
where $\alpha^\prime$ is the string tension. In weakly-coupled
string theory, $\lambda_S$ is larger than $\lambda_P$ 
and string
effects can set in and modify gravity in an intermediate regime
between the eikonal elastic scattering and black-hole 
formation~\cite{venezia}. 
These are interesting effects,
as the LHC can directly probe the nature of string 
theory~\cite{dimemp}. However, model-independent calculability 
of the elastic channel 
is lost. In this paper we will consider only the 
pure transplanckian gravitational
scattering, assuming that quantum-gravity effects are small.

The relevance of our considerations for experiments at the LHC 
and other future colliders is discussed
in sect.~3. We will compute the elastic cross section in
the transplanckian eikonal regime,
\beq
M_D/\sqrt{s}\ll 1,~~~~~-t/s\ll 1,
\label{pit}
\eeq
where $t$ is the Lorentz-invariant Mandelstam variable, and
$-t/s$ is a measure of the scattering angle in the center-of-mass
frame. The experimental signal will be
discussed and compared with the expected Standard Model (SM) 
rate and with black-hole
production. 

\section{Scattering in the Eikonal Approximation}
\label{sec2}

\subsection{The Eikonal Amplitude}
\label{sec21}

We are interested in the elastic scattering of two massless
4-dimensional
particles (living on a 3-brane) due to a 
$D$-dimensional weak gravitational
field in the kinematic limit of \eq{pit}. 
We will focus on the case in which the two colliding particles are
different, but the case of identical particles is analogous.
In the eikonal approximation,
we consider the infinite set of ladder and cross-ladder Feynman diagrams
which give the leading contribution to forward scattering.

Since we are working in the
approximation of small momentum transfer, we can compute the coupling
between the colliding particle $\Phi$ and the small fluctuation around
the flat metric (the graviton), taking $\Phi$ to be on-shell:
\beq
\frac{1}{M_D^{1+n/2}}\langle \Phi (p)|T^{\mu \nu}|\Phi (p)\rangle =
\frac{2p^\mu p^\nu}{M_D^{1+n/2}}.
\label{tmu}
\eeq
Here $T^{\mu \nu}$ is the energy-momentum tensor and $p^\mu$ is the 
4-momentum of the particle $\Phi$. Notice that \eq{tmu} is equally
valid for colliding particles of any spin.

In the $\Phi$ propagators we keep only the leading terms in the momentum
transfer $q$:
\beqa
\frac{i}{(p+q)^2} &\to& \frac{i}{2pq} ~~~{\rm for~bosons},
\nonumber \\
\frac{i}{\slash{p}+\slash{q}} &\to& \frac{i|\Phi (p)\rangle
\langle \Phi (p)|}{2pq}  ~~~{\rm for~fermions}.
\label{ferm}
\eeqa
Since the factor $\slash{p}=|\Phi (p)\rangle
\langle \Phi (p)|$ in \eq{ferm}
is used to reconstruct the on-shell vertices of \eq{tmu} in the ladder
diagrams, we observe that not only the interaction vertices but 
also the propagators of bosons and fermions 
are identical in the eikonal approximation. This universality
of bosons and fermions is not surprising since,
as discussed in the introduction, we are performing a semi-classical
calculation, and the spin of the $\Phi$ particle should not matter.

\sfig{fig1}{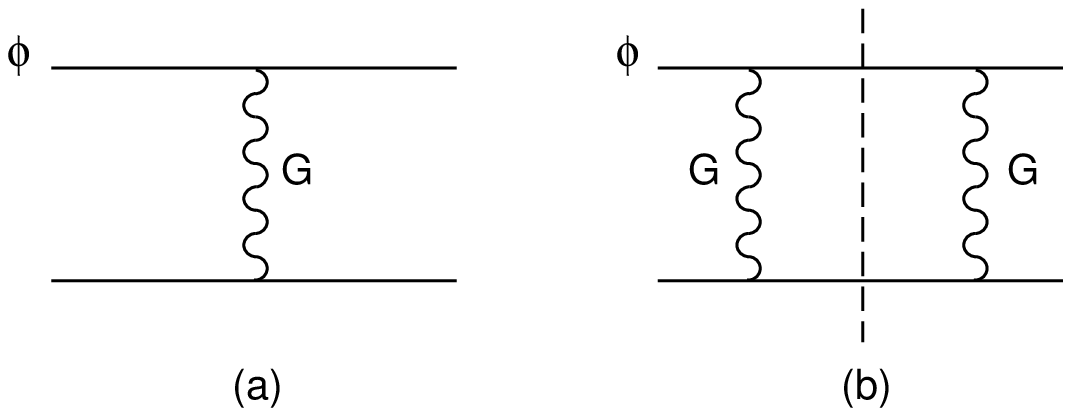}{The dominant 
Feynman diagrams contributing to elastic 
scattering of $\Phi$ particles
in the eikonal approximation at tree $(a)$ and one-loop $(b)$ 
level. Wavy lines represent the exchange of $D$-dimensional gravitons, and
the dashed line represents the cut from the physical-region
singularity.}

The tree-level exchange of a $D$-dimensional graviton in the diagram
of fig.~\ref{fig1}a gives the scattering amplitude
\beq
{\cal A}_{\rm Born} (-t)=\frac{s^2}{M_D^{n+2}} \int \frac{d^n q_T}{t-q_T^2}
=\pi^{\frac{n}{2}} \Gamma(1-n/2)\left( \frac{-t}{M_D^2}\right)^{\frac{n}{2}
-1}\left( \frac{s}{M_D^2}\right)^2,
\label{born}
\eeq
where $q_T$ is the momentum transfer in the extra dimensions. Although
\eq{born} is a tree-level amplitude, it is divergent due to the infinite
number of (extra-dimensional) momentum configurations of the exchanged
gravitons, allowed by the non-conservation of momenta transverse to the
brane. Divergences have been subtracted using dimensional regularization,
considering non-integer $n$, but the regularization prescription is
not important since the
eikonalization will consistently eliminate any ultraviolet sensitivity.
Basically this is because the eikonalization selects in \eq{born}
only the partial waves with large angular momentum. These do not depend on
the local counter-terms as they are determined by
the finite (calculable) non-analytic 
terms in $-t$, which depend only on infrared physics~\cite{noi}. These terms
are proportional to $(-t)^{(n-2)/2}\ln (-t)$ for even $n$, and to
$\sqrt{-t}(-t)^{(n-3)/2}$ for odd $n$.

Let us now turn to the one-loop amplitude. The leading term in the
limit of small momentum transfer is given by a Feynman 
diagram with a cut arising
from a physical-region singularity in which the intermediate $\Phi$
particles are on their mass shells. There is only one such diagram at one loop,
and it is drawn in fig.~\ref{fig1}b. 
All other diagrams are subleading. For instance, the one-loop $t$-channel
exchange of two gravitons $G$ coupled via quartic $GG\Phi\Phi$ vertices
is suppressed\footnote{Actually, in the spirit of
the eikonal resummation, this diagram should be compared to the
tree-level graviton exchange, but an analogous suppression factor would
nevertheless appear.} by a power of $t/s$. 
Diagrams with one $GG\Phi\Phi$ vertex and two
$G\Phi\Phi$ vertices simply vanish (as is evident by working in the
de Donder gauge).

The diagram of fig.~\ref{fig1}b with on-shell intermediate states 
can be calculated using the
Cutkosky rule. Its imaginary part is 1/2 times the discontinuity across
the cut, which is obtained by replacing the $\Phi$ propagators in
\eq{ferm} with $2\pi i \delta(2pq)$. If $p_1$ and $p_2$ are the momenta
of the incoming particles, we obtain
\bea
{\cal A}_{\rm 1-loop} (-q^2) &=&
\frac{-i}{2}\left( \frac{s^2}{M_D^{n+2}}\right)^2
\int \frac{d^4 k}{(2\pi)^4} d^n k_T d^n k_T^\prime 
\frac{1}{k^2-k_T^2}\frac{1}{(q-k)^2-k_T^{\prime 2}}
(2\pi i)^2  \delta(2p_1k)\delta(2p_2k) \nonumber \\
&=& \frac{i}{4s}\int \frac{d^2 k_\perp}{(2\pi)^2}
{\cal A}_{\rm Born}(k_\perp^2) {\cal A}_{\rm Born}[(q_\perp -k_\perp)^2].
\label{1loop}
\eea
The delta functions from the Cutkosky rule
had the effect of reducing the 4-dimensional integral
into a 2-dimensional integral over momenta perpendicular to the beam (and 
along the brane). 
Notice that, in the eikonal limit, the momentum transfer $q$ is mainly
in the perpendicular direction and therefore, we can use the approximation
\beq
t =q^2\simeq -q_\perp^2.
\eeq

Equation (\ref{1loop}) 
is merely a convolution of the Born amplitude. It is then convenient
to perform a Fourier transform of the amplitudes 
with respect to the transverse momentum $q_\perp$. This amounts to trading
$q_\perp$ with its conjugate variable $b_\perp$, the impact parameter. In
impact parameter space \eq{1loop} becomes a simple product.
Summing \eq{born} and \eq{1loop}, and transforming back to momentum space
we can recast the loop expansion in the form
\beq
{\cal A}_{\rm Born}(q_\perp^2)+{\cal A}_{\rm 1-loop} (q_\perp^2)+\dots
=-2is \int d^2b_\perp e^{iq_\perp b_\perp} \left( i\chi -\frac{1}{2}
\chi^2 + \dots \right),
\label{insie}
\eeq
\beq
\chi (b_\perp )=\frac{1}{2s} \int \frac{d^2q_\perp}{(2\pi)^2}
e^{-iq_\perp b_\perp}{\cal A}_{\rm Born}(q_\perp^2).
\label{eikp}
\eeq
The combinatorics of higher-order loop terms is such that one can resum
all terms in \eq{insie} to obtain
\beq
{\cal A}_{\rm eik} =-2is \int d^2b_\perp e^{iq_\perp b_\perp} \left( e^{i\chi}
-1\right) .
\label{eicon}
\eeq
We have recovered the known result~\cite{eik} that the eikonal scattering
phase $\chi$, function of the impact parameter plane coordinates
$b_\perp$, is the 2-dimensional Fourier transform of the Born amplitude 
in the direction perpendicular to the beam.

One crucial feature of the eikonal procedure is that $\chi(b)$ for $b\not=0$
depends only on the calculable, ultraviolet 
finite, terms in $A_{\rm Born}$. The ultraviolet
divergent terms correspond to $delta$-function singularities localized
at $b=0$. While these terms are obviously irrelevant for scattering processes
where the initial states 
are prepared in such a way that they have no overlap at $b=0$,
we will also define our full amplitude in \eq{eicon} by neglecting these terms.
We believe this to be a consistent procedure. 
Indeed, it would not make
any sense to include $\delta$-functions in the exponent of \eq{eicon}. 
From
a physical point of view, we expect these singularities to be softened
by the fundamental theory of gravity, at some finite, but small,
impact parameter $b\sim \lambda_P$
(or, more likely, $b\sim \lambda_{\rm S}$). So the presence
of these terms is just evidence that there should be corrections to our simple
eikonal expression
coming from the underlying theory of quantum gravity. We will discuss
these effects in sect.~\ref{seccor}. For the moment we limit ourselves to
note that these short-distance contributions should plausibly give rise to 
${\cal O}(\lambda_P^2)$
corrections to the cross section, while, as we will see shortly, the 
long-distance eikonal amplitude gives a cross section 
that grows with a power of $\sqrt s$,
thus dominating at large energies.  

We can now calculate explicitly the eikonal phase $\chi$ in \eq{eikp}.  
One finds (see appendix)
\beq
\chi =\frac{\pi^{\frac{n}{2}-1} \Gamma (1-n/2)s}{4M_D^{n+2}}
\int_0^\infty dq q^{n-1} J_0(qb) =\left( \frac{b_c}{b}\right)^n,
\label{bbc}
\eeq
\beq
b_c\equiv \left[ \frac{(4\pi)^{\frac{n}{2}-1}s\Gamma (n/2)}{2M_D^{n+2}}
\right]^{1/n}.
\label{defbc}
\eeq
Here $q\equiv |q_\perp |\simeq \sqrt{-t}$, $b\equiv |b_\perp |$ and
$J_0$ is the zero-th order Bessel function. 
Notice that by inserting this result for $\chi$ in the integral in
\eq{eicon} we obtain an ultraviolet 
finite result. This is so, even though the contributions to \eq{eicon}
from the individual terms
in the expansion of $e^{i\chi}=1+i\chi+\dots$ are ultraviolet 
divergent, corresponding
to the fact that each individual Feynman diagram in the ladder expansion is 
ultraviolet
divergent but the complete sum is finite. 
Moreover, since $\chi\propto b^{-n}$, the integrand in \eq{eicon}
oscillates very rapidly as $b\to 0$, showing that the ultraviolet 
region gives but
a small contribution to the amplitude.

\begin{figure}[tpb]
\dofig{6in}{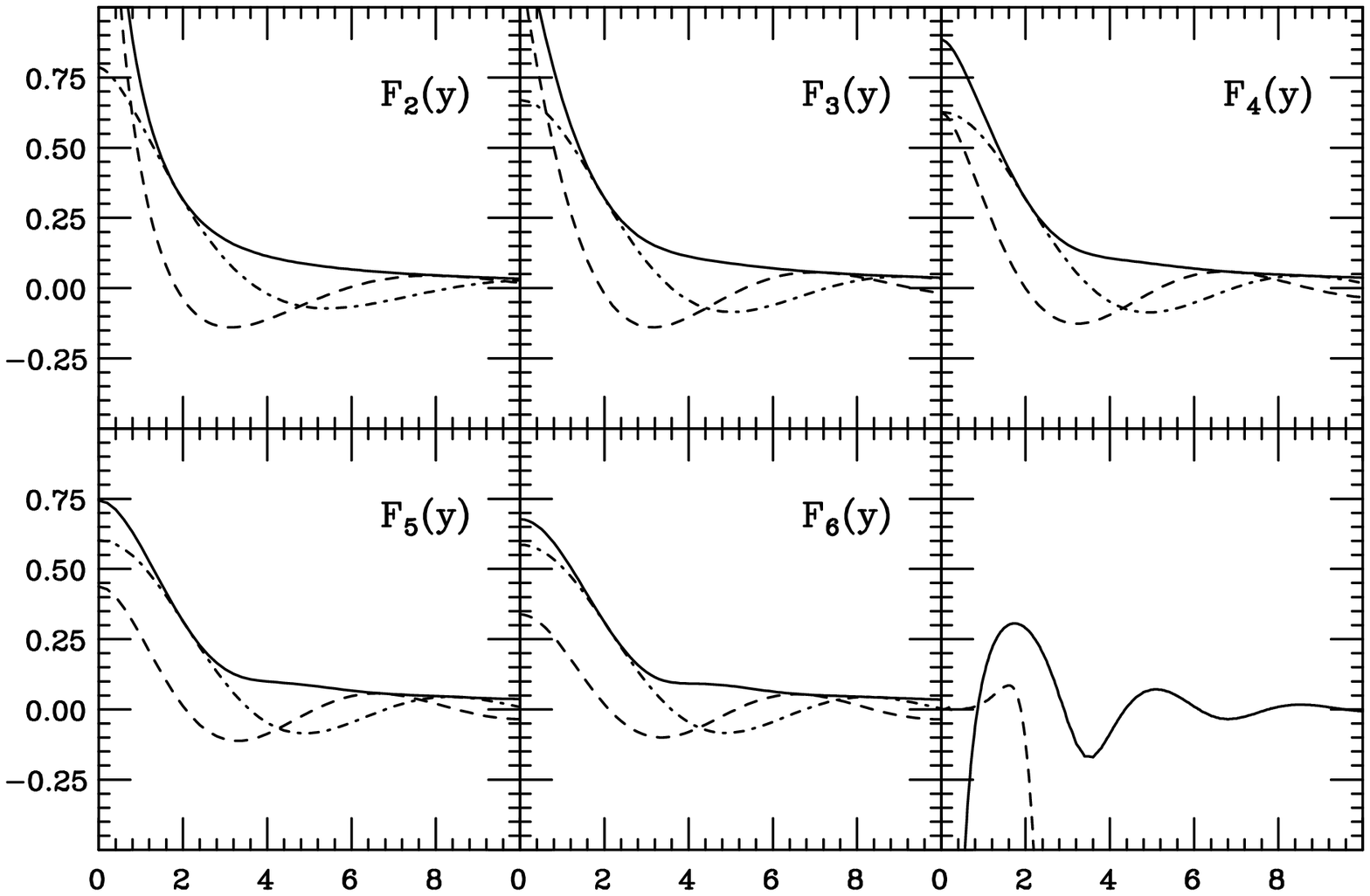} 
\caption{Plots of the function $F_n(y)$ versus $y$ for $n$ between 2 and 6.
The dashed line is the real part, the dash-dotted
line is the imaginary part, and the solid line is the absolute
value of the function. The bottom right panel plots the
relative error of $|F_6(y)|$ compared to the asymptotic expressions,
see Eqs.~(\ref{singg}) and~(\ref{statp}),
as $y\to 0$ (dashed line) and $y\to\infty$ (solid line).}
\label{fig2}
\end{figure}

Replacing \eq{bbc} into \eq{eicon}, we obtain the final expression 
for the eikonal amplitude~\cite{ratta}
\beq
{\cal A}_{\rm eik} =4\pi s b_c^2 F_n (b_c q),
\label{eikfin}
\eeq
\beq
F_n (y)=-i\int_0^\infty dx x J_0 (xy) \left( e^{ix^{-n}}-1\right) ,
\label{funn}
\eeq
where the integration variable is $x=b/b_c$.
The functions $F_n$, which can be written in terms of Meijer's G-functions,
are shown in fig.~\ref{fig2}.  

At small $y$, the functions $F_n (y)$ can be expanded as (see appendix)
\bea
F_2(y)&\stackrel{\textstyle =}{\scriptstyle y\to 0}&-
\ln \frac{y}{1.4} +i\frac{\pi}{4} \\
F_3(y)&\stackrel{\textstyle =}{\scriptstyle y\to 0}
&\frac{i}{2}\Gamma \left( \frac{1}{3} \right) 
e^{i\frac{\pi}{3}} -y\\
F_4(y)&\stackrel{\textstyle =}{\scriptstyle y\to 0}
&i\frac{\sqrt{\pi}}{2}
e^{i\frac{\pi}{4}}+\frac{y^2}{4}\ln y\\
F_n(y)&\stackrel{\textstyle =}{\scriptstyle y\to 0}
&\frac{i}{2}
\Gamma \left( 1-\frac{2}{n} \right) e^{-i\frac{\pi}{n}}
-\frac{i}{16}y^2\Gamma \left( 1-\frac{4}{n} \right) e^{-i\frac{2\pi}{n}}
~~~{\rm for~}n>4. \label{singg}
\eea
These functions develop non-analytic terms in $y^2$ of the form $y^{n-2}\ln y$
for even $n$, and $y^{n-2}$ for odd $n$, which reproduce the non-analytic
structure of ${\cal A}_{\rm Born}$ previously mentioned.
Notice that $F_n (0)$ is finite for $n>2$.

For $y\gg 1$ ($q\gg b_c^{-1}$), the stationary-phase approximation applies, yielding
\beq
F_n(y)\stackrel{\textstyle =}{\scriptstyle 
y \gg 1} \frac{-in^{\frac{1}{n+1}}y^{-\frac{n+2}{n+1}}}
{\sqrt{n+1}} \exp \left[ -i(n+1)\left( \frac{y}{n}\right)^{\frac{n}{n+1}}
\right] .
\label{statp}
\eeq
The function $F_n$ oscillates around its asymptotic value given in
\eq{statp}, before converging to it at large $y$, as illustrated in the
bottom-right panel of fig.~\ref{fig2}.

\subsection{Cross Sections}

Let us now consider the differential cross section for elastic scattering.
From \eq{eikfin} we obtain
\beq
\frac{d\sigma_{\rm eik}}{dt}=\pi b_c^4 \left| F_n (b_c\sqrt{-t})\right|^2.
\label{sigpr}
\eeq
We recall the definitions
\beq
R_S^2s=A_n \left(\frac{s}{M_D^2}
\right)^{\frac{n+2}{n+1}},~~~~~ b_c^2s=B_n \left(\frac{s}{M_D^2}
\right)^{\frac{n+2}{n}},
\label{defab}
\eeq
where the coefficients $A_n$ and $B_n$ are given in table~1.
In terms of the geometric black-hole cross section
$\sigma_{BH}\equiv \pi R_S^2$, \eq{sigpr} can be written as
\bea
\frac{d\sigma_{\rm eik}}{dt}&=&
\frac{B_n^{\frac{n}{n+1}}}{A_n} 
\frac{\sigma_{BH}}{s} \left( b_c^2 s \right)^{\frac{n+2}{n+1}}
\left| F_n (b_c\sqrt{-t})\right|^2 \nonumber \\
&=& \frac{B_n^2}{A_n} 
\frac{\sigma_{BH}}{s} \left(\frac{s}{M_D^2}
\right)^{\frac{(n+2)^2}{n(n+1)}} \left| F_n (b_c\sqrt{-t})\right|^2 .
\eea

\begin{table}
\label{tab1}
\centering
\begin{tabular}{|c|c||c|c|c|c|c|}
\hline\hline
 & $n=$ & 2 & 3 & 4 & 5 & 6 \\
\hline
$A_n=$ & $\left[ \frac{2^n  \pi^{\frac{n-3}{2}}
 \Gamma \left( \frac{n+3}{2}\right)}{n+2} \right]^{\frac{2}{n+1}}$
& 0.83 & 1.79 & 3.01  & 4.42 & 5.96  \\
$B_n=$ & $\left[ 2^{n-3}\pi^{\frac{n-2}{2}}\Gamma \left( \frac{n}{2}\right)
\right]^{\frac{2}{n}}$  
& 0.50 & 1.35  &2.51  &3.88  & 5.41 \\
\hline\hline
\end{tabular}
\caption{{The coefficients $A_n$ and $B_n$ entering \eq{defab} and their
numerical values for $n$ between 2 and 6.
}}
\end{table}

In the limit $q\ll b_c^{-1}$, or equivalently
$-t/s \ll B_n^{-1} (M_D^2/s)^{1+2/n}$, the
differential cross section becomes (for $n>2$)
\bea
\frac{d\sigma_{\rm eik}}{dt}&=&
\frac{[\Gamma (1-2/n)]^2B_n^{\frac{n}{n+1}}}{4A_n} 
\frac{\sigma_{BH}}{s} \left( b_c^2 s \right)^{\frac{n+2}{n+1}}
\nonumber \\
&=& \frac{[\Gamma (1-2/n)]^2B_n^2}{4A_n} 
\frac{\sigma_{BH}}{s} \left(\frac{s}{M_D^2}
\right)^{\frac{(n+2)^2}{n(n+1)}} .
\label{diff1}
\eea
In the opposite limit ($q\gg b_c^{-1}$), we find
\beq
\frac{d\sigma_{\rm eik}}{dt}= \frac{\left( n^2B_n^n\right)^{\frac{1}{n+1}}}{
(n+1)A_n} \frac{\sigma_{BH}}{s}
\left( -\frac{s}{t}\right)^{\frac{n+2}{n+1}}.
\label{diff2}
\eeq
Notice how the two formul{\ae} valid in the two regimes match since, for
$-t\simeq b_c^{-2}$, \eq{diff2} parametrically agrees with \eq{diff1}.
Moreover, extrapolating \eq{diff2} beyond its range of validity 
towards small impact parameters ({\it i.e.} large angle, or $t \to -s$),
we find a result that matches parametrically the black-hole production cross 
section derived from dimensional analysis.

The integrated elastic cross can be derived directly from \eq{eicon}:
\bea
\sigma_{\rm el}&=& \frac{1}{16\pi^2s^2}\int d^2q_\perp
\left| {\cal A}_{\rm eik} \right|^2 \nonumber \\
&=& \int d^2 b_\perp \left( 1+e^{-2{\rm Im}\chi}
-2e^{-{\rm Im}\chi} \cos {\rm Re} \chi \right) .
\label{sigtot}
\eea
The total cross section, inclusive of elastic and inelastic
channels, can be derived from the optical theorem
\beq
\sigma_{\rm tot} =\frac{{\rm Im}{\cal A}_{\rm eik} (0)}{s}
=2 \int d^2 b_\perp \left( 1-e^{-{\rm Im}\chi} \cos {\rm Re} \chi \right).
\label{sigop}
\eeq
In the absence of absorptive parts (${\rm Im}\chi =0$) \eq{sigtot}
and \eq{sigop} coincide.
Black-hole production may be modelled by a large absorptive part 
(${\rm Im}\chi \gg 1$) at impact parameters smaller than $R_S$. 

\subsection{Physical Interpretation}
\label{physical}

Before proceeding to the phenomenological analysis, we remark
on the physical meaning of the expression of the
eikonal amplitude given by \eq{eicon}
and \eq{bbc}. We can first consider a non-relativistic analogue of the
problem under investigation~\cite{landau} by studying
the scattering of a particle
with mass $m$, velocity $v$, and impact parameter $b$ 
by a radial ``gravitational'' potential
\beq
V(r)=\frac{G_D mM}{ r^{n+1}}.
\eeq
Here $m$ and $M$ are the non-relativistic analogues of $\sqrt{s}/(2c^2)$,
$v$ is the analogue of $c$, and $n$ corresponds to the number of extra
dimensions. 

When $b$ is not too small, the quantum-mechanical scattering
phase of the wave associated with angular momentum $\ell \hbar =mvb$ is
\beq
\delta_b =-\frac{1}{v\hbar} \int_b^\infty dr V(r).
\label{delsc}
\eeq
The intergral in \eq{delsc} converges only for $n>0$, leading to
$\delta_b =-(b_c/b)^n$, where $b_c\simeq [G_DmM/(\hbar v)]^{1/n}$ exactly
corresponds to the definition in \eq{defbc}.
Therefore, the forward amplitude (which is proportional to
$\int db~ b\delta_b$) is finite only for $n>2$. For $n=0$ one finds
the Coulomb singularity, and for $n=2$ a logarithmic divergence.
This is in agreement with \eq{singg} and with the relativistic results
presented in sect.~\ref{sec21}.

A classical description of the process is valid if the 
quantum-mechanical uncertainties
in the impact parameter $b$ and in the scattering angle $\theta$
are small with respect
to their classical values. Using the Heisenberg principle, the quantum
uncertainties are estimated to be 
\beq
\Delta \theta \sim \frac{\Delta q}{mv} \sim \frac{\hbar}{mv\Delta b},
\eeq
where $q$ is the momentum in the direction orthogonal to the initial
velocity $\vec v$.
The classical scattering angle is equal to the force ($dV/dr$) at
the minimum distance ($r=b$), times the collision time ($b/v$),
divided by the momentum ($mv$),
\beq
\theta \sim \frac{b}{mv^2}
~\frac{dV(b)}{db}.
\eeq  
The condition for the validity of the classical approximation 
($\Delta \theta \ll \theta$, $\Delta
b \ll b$) implies $|dV(b)/db| >\hbar v/b^2$.
For a Coulomb-like potential $V(r)=\alpha /r$, this condition is satisfied
in the non-perturbative region $\alpha > \hbar v$. In our case, it implies
(for $n>0$) $b<b_c$, with $b_c \sim [G_D mM/(v\hbar )]^{1/n}$.
The length scale $b_c$ is a quantum scale ($b_c\to \infty$ as $\hbar \to 0$)
characteristic of the extra-dimensional 
scattering (it is undefined at $n=0$). Therefore, at $b>b_c$, quantum 
mechanical  effects
cannot be neglected. 

In the classical process (for $n>-1$), a projectile with an arbitrarily
large impact parameter $b$ is deflected by a non-vanishing (although
very small) angle, and therefore the cross section diverges at
$q=0$ ($b\to \infty$). In our case, for $b>b_c$, quantum mechanics
sets in, and the quantum fluctuations of the scattering angle become
larger than its classical value, rendering ill-defined the notion of
a trajectory. This is why the quantum scattering amplitude can remain
finite at $q\to 0$.

It is also interesting to compare the result of \eq{eikfin} with the
fully $D$-dimensional case, in which both gravity and the colliding
particles propagate in the extra dimensions~\cite{venezia}. In this case
\eq{eicon} becomes
\beq
{\cal A}_{\rm eik} =-2is \int d^{2+n}b_\perp e^{iq_\perp b_\perp} \left[ 
e^{i(b_c/b_\perp)^n}
-1\right] .
\label{eqsuper}
\eeq 
For $q=0$, the integral is infrared
dominated by large values of $b$ and it is
quadratically divergent (for any $n$). Therefore we find ${\cal A}_{\rm eik}
(q\to 0) \sim b_c^n /q^2$, and we encounter the Coulomb singularity
characteristic of long-range forces. On the other hand, in the case
of particles living on the 3-brane, we have found (for $n>2$) 
that the forward scattering amplitude is finite and that, although the force
does not have a finite range, the integral in \eq{eicon} at $q=0$ is
dominated by distances of order $b_c$. Indeed, by
working with localized particles we lose momentum conservation 
in the transverse direction,
and moreover the Born amplitude becomes dominated by contact terms of 
ultraviolet origin. 
The fact that \eq{eicon} is not infrared dominated even for $q\to 0$ is a  
remnant of this property of the Born amplitude.

Let us now come back to the relativistic elastic scattering under 
consideration.
As we have seen in sect.~\ref{sec21}, there exist two 
different kinematic regions
in which the eikonal amplitude has distinct behavior.

Region {\it (i)}:~~~ $\sqrt{s}\gg q \gg b_c^{-1}$.\\
This is the region in which the integral in \eq{funn} is dominated by the
stationary-phase value of the impact parameter $b=b_s$, where
\beq
b_s\equiv
b_c \left( \frac{n}{qb_c}\right)^{\frac{1}{n+1}} =
\frac{\left( nB_n^{n/2}\right)^{\frac{1}{n+1}}}{\sqrt{s}} \left(- \frac{t}{s}
\right)^{-\frac{1}{2(n+1)}} \left( \frac{s}{M_D^2}\right)^{\frac{n+2}{2(n+1)}}.
\label{defbs}
\eeq
In this region $b<b_c$, and
therefore the eikonal phase $\chi=(b_c/b)^n$ is large
(in units of $\hbar$): we are in the classical domain. 
Indeed, the classical scattering angle is given by the derivative
of the scattering phase with respect to the angular momentum $L=\sqrt{s}
b/2$,
\beq
\theta_{\rm cl}=-\frac{\partial \chi}{\partial L}=\frac{2n\Gamma (n/2)}{
\pi^{n/2}}~\frac{G_D\sqrt{s}}{b^{n+1}}.
\label{einsa}
\eeq
In the limit $n\to 0$, we recover the Einstein angle $\theta_{\rm cl}
=4G_D\sqrt{s}/b$
 while, for $n>0$, \eq{einsa} gives its higher-dimensional
generalization\footnote{The Einstein angle defines the deflection of a photon
by the static gravitational field of a mass-$M$ particle at rest: 
$\theta_E=4G_NM/b$. It is easy to see that by boosting $\theta_E$
to the center-of-mass frame, in the limit 
$\sqrt s \gg M$, we obtain our expression of $\theta_{\rm cl}$.}. 
Notice that the relation between scattering angle and
impact parameter in \eq{einsa} is equivalent to $b=b_s$.

Region {\it (ii)}:~~~ $ q \ll b_c^{-1}$. \\
In this region the integral in \eq{funn} is dominated by $b$ of the
order of (or slightly smaller than) $b_c$. This means that the eikonal
phase $\chi=(b_c/b)^n$ is of order one 
(in units of $\hbar$) and the quantum
nature of the scattering particles is important (although the exchanged graviton
is treated as a classical field, and so quantum gravity effects are 
negligible), as we could have expected from the discussion of the
non-relativistic analogue presented above.
There is no one-to-one relation between impact parameter and scattering
angle in region {\it (ii)}, since the classical concept of a trajectory
is not applicable. 
Moreover, notice that the relevant $\chi$ never becomes much smaller than 1,
and therefore we never enter the perturbative regime in which a loop
expansion for the amplitude applies. Even though the interaction vanishes
at $b\to \infty$ (where $\chi \to 0$), we never reach the Born limit.
Even for $q=0$, the scattering is dominated by $b=b_c$ and not 
by $b= \infty$, as in the Coulomb case. As explained after \eq{eqsuper},
this result simply follows from the
different dimensionalities of the spaces relevant for
scattering particles (living on a 3-brane)
and exchanged graviton (propagating in the bulk), and it does not hold
in the case of scattering of bulk particles.

\subsection{Corrections to the Eikonal Approximation}
\label{seccoreik}

The conditions required by 
our approximations are expressed in \eq{pit}. In order
to assess the theoretical uncertainty inherent to our calculation, in
this section we estimate the size of the effects that we have
neglected. 
We first discuss classical
effects, for which general relativity suffices. We also include the
quantum-mechanical effects of region {\it (ii)}, but we leave quantum-gravity
contributions for the next section.
The neglected classical terms come from the approximations
of small angle ($-t/s \ll 1$) and of weak gravitational field
($R_S/b \ll 1$).
Associated with the subleading classical effects there is also the emission of
gravitational 
radiation, leading to missing energy signals at high-energy colliders.

Let us estimate these effects. 
 Notice that in the classical region {\it (i)}
where $\sqrt{s}\gg q\gg b_c^{-1}$, the two 
requirements coincide since $b=b_s$, see \eq{defbs}. 
Here, scattering can be
described in terms of
classical trajectories. The equation $b=b_s$ represents a relation
between transferred momentum and impact parameter 
\beq
-\frac{t}{s}\sim \frac{G_D^2 s}{b^{2n+2}}\sim \left( \frac{R_S}{b}
\right)^{2n+2},
\label{leading}
\eeq
where we have neglected factors of order unity. 
This result represents an approximation~\cite{hoof}
in which one considers only a linear superposition of the
gravitational shockwave fields generated by the two colliding 
particles~\cite{Aichelburg}, thereby
neglecting non-linear effects of their mutual
interactions. In this computation, each field
represents an exact solution to Einstein's equations
in the absence of the other field.
Inclusion of the non-linear effects will induce  ${\cal O}(G_D^p)$ corrections
to \eq{leading}, where $p$ is an unknown power. 
By dimensional analysis the relative size of
these corrections must be  
$G_D^p s^{p/2}/b^{p(n+1)} \sim (R_S/b)^{p(n+1)}$.
Moreover since we expect that the evaluation of these effects is perfectly
perturbative, analyticity in $s$ should hold, and therefore only even integers
of the exponent $p$ will appear. We conclude that \eq{leading} will be 
corrected by a factor of the form
\beq
1+{\cal O}\left( \frac{G_D^2 s}{b^{2n+2}}\right) =1+{\cal O}\left( 
\frac{t}{s}\right) .
\label{subleading}
\eeq
This will induce a correction of the same size to the
cross section, in the classical region {\it (i)}. Notice that \eq{subleading}
corresponds to a ${\cal O}(\theta^3)$ correction term
to the relation between scattering angle and
impact parameter, \eq{einsa}. In the scattering 
by a static potential we would get corrections
already at ${\cal O}(\theta^2)$. Our different 
result follows from the relativistic
nature of the process.

In region {\it (ii)} where $q\ll b_c^{-1}$,
the leading correction to the cross section will then have the form
\beq
\frac{d \sigma}{ d t}= \frac{d \sigma_{\rm eik}}{ d t}\left[ 1
+ \frac{t}{s}\, Z(tb_c^2)\right ] ,
\label{fullcorr}
\eeq
where $Z$ is expected to be a constant of order unity in the classical 
region $t b_c^2\gg 1$.
In the region $t b_c^2\ll 1$, the quantum nature of the scattering particles
becomes relevant, and $Z$ will have non-trivial behavior. However,
admitting only finite corrections at $t\to 0$ implies that
the size of the relative correction to $d\sigma/dt$ is at most
$1/sb_c^2\sim (R_S/b_c)^{2n+2}$.

In summary, we have inferred that the leading classical gravity
corrections  to 
the eikonal amplitude are 
\beq
{\cal O}\left( -\frac{t}{s}\right) +{\cal O} 
    \left[ \left( \frac{M_D^2}{s}\right)^{1+\frac{2}{n}} \right] ,
\label{correx}
\eeq
and therefore suppressed and under control for the conditions we
are applying, as expressed by \eq{pit}.

The discussion outlined above is consistent with the analysis of 
ref.~\cite{venezia} where classical corrections to the eikonal are identified
by resumming an improved ladder series including  a class of
two-loop graphs, the so-called H-diagrams. After improving the Born
term by the H-diagram, the eikonal phase is modified to
\beq
\chi \quad \to \quad \left( \frac{b_c}{b}\right)^n 
\left [1+ {\cal O}\left(\frac{R_S}{b}\right)^{2n+2}
\right] .
\label{quarantaq}
\eeq 
It is easy to check that altering \eq{eicon} according to 
the above equation produces corrections to the leading amplitude
consistent with \eq{correx}.
Notice that the corrections to the eikonal become large for impact parameters
comparable to the Schwarzschild radius, for which the production
of black holes presumably sets in\footnote{The hypothesis that classical
physics determines the black-hole production cross section has
been criticised in ref.~\cite{voloshin}. However, we believe that
the appearence of
corrections of the form shown in \eq{quarantaq} is a further indication
that non-trivial classical dynamics emerges at $b\simeq R_S$.}. 

Gravitational radiation is associated
with subleading classical effects. Emission of gravitons  is signaled
by the presence of an imaginary part in the H-diagram  
contribution to the eikonal phase \cite{venezia}
\bea
{\rm Im} (\chi_H) &\sim &\left (\frac{b_r}{b}\right )^{3n+2},
\label{imclass}\\
b_r &\equiv &\left(b_c^n R_s^{2n+2}\right)^{\frac{1}{3n+2}}\sim
\left( G_D^3 s^2 \right)^{\frac{1}{n+2}}. 
\nonumber
\eea
This absorptive term corresponds to a depletion of purely elastic scattering
due to the emission of gravitational radiation.
When $b\ll b_r$ the probability of elastic scattering becomes
exponentially small. The calculation of ref. \cite{venezia} shows that the
emitted radiation has typical transverse momentum $\sim 1/b$.
However the longitudinal momentum is distributed up to values of order
$\sqrt s$.
Therefore we expect that for $b<b_r$ a significant fraction of the energy
is radiated in the form of forward gravitational radiation.
Notice  that $b<b_r$ is typically outside our chosen kinematical regime,
so that we expect only a fraction $\sim (b_r/b)^{3n+2}\ll 1$ of initial
energy to be lost to invisible gravitational radiation.
In the semiclassical region, we find $(b_r/b)^{3n+2}\simeq (-t/s)^{n/(n+1)}
(\sqrt{-t}/M_D)^{(n+2)/(n+1)}$, and therefore gravitational radiation for
small-angle scatterings is not large.

One may worry that important emission of radiation, already at $b\gg R_s$,
may drastically reduce the naive geometric estimate of the black-hole 
production cross-section. Indeed if one simply interpreted  
\eq{imclass}, {\it \`a la} Block-Nordsiek, as the number $N$ of emitted 
gravitons,
one would conclude that hard bremsstrahlung depletes
all the energy, as soon as $b$ is smaller than $b_r$. 
However this naive interpretation
is manifestly inconsistent as the total emitted energy would quickly exceed
$\sqrt s$. Indeed it is likely that the exponentiation of the H-diagram
imaginary part is inconsistent for hard radiation. On the other hand
for soft radiation with $E\sim 1/b$, the Block-Nordsiek interpretation
is likely to be reasonable. Thus if the number of soft gravitons
is $N_{\rm soft}\sim (b_r/b)^{3n+2}$ 
then the total energy lost to soft radiation
at large angle is
\beq
E_{\rm rad}\sim \frac{N_{\rm soft}}{b}\sim {\sqrt s}\left (
\frac{R_S}{b}\right )^{3n+3},
\eeq
which becomes important only for $b\sim R_S$.
 
\subsection{Quantum-Gravity Effects}
\label{seccor}

On general grounds we expect that quantum-gravity contributions
to the eikonal phase must be of the form  $\delta \chi/\chi\sim (
\lambda_P/b)^k$,
with $k>0$ and $\lambda_P$ given in \eq{lamp}.
If we work in the true transplanckian region ($\sqrt{s}\gg M_D$), these
effects are subleading\footnote{It is interesting to compare the case 
of gravity with the case of vector-boson exchange which 
has been mentioned in sect.~1. For a vector, the
eikonal phase is $\chi_V\propto e^2/b^n=1/(M_eb)^n$ (we take $\hbar=c=1$).
Notice that, as opposed to gravity, there is no factor of $s$ 
enhancement~\cite{kab1}. It is easy to
realize that quantum corrections will also go like a power of $e^2/b^n$. 
This is
for instance the case for vacuum-polarization improvement of the vector
propagator. Therefore, there is no impact parameter choice for which
both $\chi_V \gg 1$ and quantum corrections are small, where the eikonal
approximation can be useful.}  
 since, as we have discussed in sect.~1, the
classical scale $R_S$ is much larger than the quantum scale $\lambda_P$.

However, as it will be clear in the following, collisions at the LHC
can only barely satisfy the transplanckianity condition.
Therefore, quantum-gravity
corrections are potentially large, and they are likely to be the
limiting factor of our approximations. This happens because, although
quantum-gravity effects are characterized by a length scale $\lambda_P$
smaller than the \sch radius $R_S$, classical corrections appear with
a large exponent, see eqs.~(\ref{subleading}) and (\ref{correx}), 
remnant of the geometric nature
of the interaction, while this is not expected to be the case 
for the quantum-gravity corrections ${\cal O}(\lambda_P/b)^k$.
 
However in order to be able to say more we need a quantum-gravity
model or at least
a framework. String theory is the only possibility at hand, and the
transplanckian regime has been discussed in this context 
by several authors, see {\it e.g.} refs.~\cite{venezia,sold,grossmende}.  
Indeed, we will show that string corrections
appear to be large for energies relevant to LHC experiments.

In ref.~\cite{venezia} string corrections to eikonalized graviton scattering
for type II superstrings were studied. Since we do not stick to a particular
string realization of our brane-world scenario, we will limit ourselves
to a qualitative discussion. 
Following ref.~\cite{aadd} we may imagine a realization of the 
brane-world in type I,
where $m=6-n$ spatial dimensions out of the 10 space-time dimensions
are compactified
on a torus $T^m$ with radius $r_m={\sqrt {\alpha'}}=1/M_S$, and $M_S$ is the 
string scale. Of the remaining dimensions, $n$ 
have a ``large'' compactification radius and 3 are the
usual non-compact ones.
 The brane-world is realized by a $(3+m)$-dimensional D-brane, in 
which $m$ dimensions
span the small manifold $T^m$. In terms of $M_S$ and of the string coupling 
$g_S$,
the $4+n$ dimensional Planck scale $M_D$ and the gauge coupling $g$
on the brane are given by
\beq
M_D^{2+n}=\frac{e^{-2\phi}}{\pi}M_S^{2+n},\quad\quad\quad \frac{1}{g^2}=
\frac{e^{-\phi}}{2\pi},\quad\quad\quad g_S=e^\phi ,
\label{ancora}
\eeq
where $\phi$ is the dilaton vacuum expectation value. Of course these
relations will receive corrections of order unity in realistic models
with broken supersymmetry and should be viewed, here and in the following,
only as indicative estimates. 
Equations~(\ref{ancora}) imply
\beq
\left (\frac{M_S}{M_D}\right )^{2+n}=\frac {g^4}{4\pi}.
\eeq
If we identify $g$ with the gauge couplings of the SM, 
say for instance $g=g_{\rm weak}$, we conclude that there should be a mild 
hierarchy
between $M_D$ and $M_S$. For instance, we find $M_D/M_S$ equals 1.7 
and 2.8 for $n=6$ and
$n=2$, respectively. {\it A priori} one could also conceive string models 
where $M_D$ and $M_S$ coincide,
with a self-dual dilaton vacuum expectation value  
$\phi\sim 0$, corresponding to a strongly coupled string.
 In this case
the weakness of the SM gauge couplings would have to be explained by 
some additional mechanism. 

As we will discuss in sect.~\ref{sec3}, in order to observe transplanckian
scatterings at the LHC, we have to consider values of $M_D$ smaller than a
few TeV. The existence of string theory at such a low scale is already
limited by precision electroweak data and by the non-observation of
new contact interactions at LEP. Exchanges of gravitons and massive string
states give rise to effective operators involving SM fields.
The leading effects come from 4-fermion operators of dimension 6 which, 
if present at tree level, 
give an approximate bound $M_S>3 $~TeV~\cite{benakli}.
This bound is in conflict with the working assumptions necessary to have
observable transplanckian signals at the LHC, and therefore we have to assume
a mild suppression of the dimension-6 operators. 
This assumption is probably not unreasonable, as there exist
examples \cite{pesk,benakli} where these effects vanish at tree level. 
In this respect the 
presence of supersymmetry at some stage in the construction may help
suppress these contributions.
Loop effects are within current experimental bounds even for $M_D$ 
of order a few TeV,
if the underlying quantum-gravity theory does not become strongly-interacting
or, in a more operative sense, if the divergent loops are cut off at a 
scale slightly smaller than $M_D$~\cite{rattas}. This allows for the
weakly-interacting string scenario in which there is a small hierarchy
between $M_S$ and $M_D$, but it essentially rules out the case of
strongly-coupled strings with $M_S$ of a few TeV, where
we would need the rather strong, and implausible, assumption that 
dimension-6 operators
are suppressed to all orders in perturbation theory. 

With these limitations in mind we now consider 
string corrections~\cite{venezia}.
The basic result is that the eikonal elastic field-theory amplitude
gets promoted to a unitary matrix acting non-trivially over a subspace of  
the string Fock space 
\bea
& &\exp \left[i \left(\frac{b_c}{b} \right)^n \right]\quad \to  \quad 
\exp \left[i \left(\frac{b_c}{b} \right)^n \right] K, 
\label{diffractive}\\
& &K = \exp \left[i \left(\frac{b_D}{b} \right)^{n+2} \hat H
\right], ~~~~b_D^{n+2}\equiv
\alpha' b_c^n . \nonumber
\eea
Here $\hat H$ is a hermitian operator involving the creation and annihilation
operators for the string oscillator modes. The above equation 
shows that,
for $b<b_D$, a diffractive production of excited string
modes takes place without any suppression. Basically what happens is that the
two colliding partons are excited into string modes through multiple soft 
graviton exchange.

The average mass of the string modes is roughly given by $M_S (b_D/b)^{n+2}$. 
This is smaller
than the available energy $\sqrt s$ as long as $(R_S/b)^{n+1}({\sqrt 
{\alpha'}}/b)<1$.
This condition is clearly satisfied within our chosen kinematical regime, 
and therefore the
produced particles are relativistic. The opening of an (exponentially)
large number of new channels drastically
depletes the elastic one. For instance in the example discussed in 
ref.~\cite{venezia},
where gravitons are scattered in type II superstring, the elastic amplitude 
correction factor becomes
\beq
\langle K\rangle =\Gamma^{2n+2}(1-i\Delta)
\Gamma^2[1-i(n+1)\Delta],
\label{elastdiff}
\eeq
where $\Delta=(n/2)(b_D/b)^{n+2}$ and where by $\langle K\rangle$ we 
indicate the expectation value of the operator $K$ on the initial state. 

For $b\gg b_D$, the correction factor in \eq{elastdiff} becomes
\beq
\langle K\rangle \simeq 1+i~2~\gamma ~ 
n(n+1) \left( \frac{b_D}{b}\right)^{n+2},
\label{corrinf}
\eeq
where $\gamma=0.577$ is the Euler number. Therefore, the effects of string 
excitations set in much before the string scale. Already at $b\sim b_c$, 
the terms modifying the gravitational eikonal phase implied by
\eq{corrinf} are as
large as $n(n+1)(M_D^2/M_S^2)(M_D^2/s)^{2/n}$. 
As we have previously anticipated,
these effects can be much more sizable than the classical corrections
in \eq{correx}, even if the string length is smaller than the classical
\sch radius.

For $b\ll b_D$, we find
\beq
|K| \simeq (n+1) (\pi n)^{n+2}\left( \frac{b_D}{b}\right)^{(n+2)^2}
\exp \left[ -\pi n(n+1)\left( \frac{b_D}{b}\right)^{(n+2)}\right],
\eeq
showing that the absolute value of the correction factor
decreases exponentially at $b<b_D$, because of the appearence of
the string excitations. 
 
This string effect is fairly general and we expect it in realistic
models not to differ too much from \eq{elastdiff}. However the sharp drop in
the elastic cross section can be compensated by looking at the inclusive diffractive
cross section. Since $b_D<b_c$ we can use the saddle point approximation to
evaluate the amplitude in transverse momentum space
\beq
{\cal A }(q)= {\cal A}_{\rm eik}(q) \, \exp \left[ i \left(\frac{b_D}{b_s(q)}
\right)^{n+2} \hat H\right] ,
\label{unitary}
\eeq
where ${\cal A}_{\rm eik}$ is the eikonal amplitude in the 
stationary phase regime.
Due to unitarity of the second factor, the total cross-section obtained by
summing  on the diffractive channels is just the same as in the absence 
of strings,
{\it i.e.} there is no exponential suppression.
Notice that this inclusive cross section is consistently defined  in 
the limit in which 
the string modes are relativistic:
at a fixed $q$ all modes come out at essentially the same angle $\theta$.
The question of how well one can measure experimentally the inclusive 
cross section
depends very much on the model, and in particular on the decay 
properties of the
produced string modes. These states are excited by gravitons, so that they
have the same color and electric charge of the incoming partons. 
If they are stable, the final state of the collision is given by
two (massive) jet events. If they are unstable and 
their dominant decay mode involves bulk states, like the graviton, 
then the presence of missing energy
would make it impossible to deduce the original center-of-mass 
energy and to make sure
that only events at a given center-of-mass energy are selected. On 
the other hand if they dominantly
decay on the brane this reconstruction might be possible. Notice that 
the coupling to
bulk modes (closed strings) and to brane modes (open strings) are 
proportional to 
$g_S^2$ and $g_S$, respectively. Therefore the decay to brane modes, 
when possible, dominates over bulk decays in weakly-coupled string models.

We also want to comment on the string effects arising in the regime
$M_D<\sqrt s<M_S/g_S^2$~\cite{dimemp}. In this regime, although the energy
is
transplanckian, $\lambda_S>R_S$ holds. Then the large angle, small impact
parameter effects are dominated by string physics instead of classical
gravitational effects. In this regime one should
use the string version of the eikonal phase~\cite{venezia}. The main
effect
is that ${\rm Im}\chi\sim g_S^2 s\alpha'\exp (-b^2/2\alpha'\ln s\alpha')$,
corresponding to inelastic  production of very excited strings (string
balls)
dominating the scattering for  $b \leq \lambda_S$. The corresponding cross
section
is $\sim \alpha'$ for $M_S/g_S<\sqrt s <M_S/g_S^2$. At $\sqrt s\sim
M_S/g_S^2$,
we have that $\alpha'\sim R_S^2$, consistent with the string balls
becoming
indinguishable from ordinary black holes~\cite{dimemp}.
The onset of inelastic string production marks the end of the validity
of our description,
since any reminiscence of classical gravitational dynamics is lost.

Notice that in the context of strings there exists another approach
to calculate fixed-angle scattering in the regime $\sqrt s\gg M_S$
\cite{grossmende}.
This different approach leads to an elastic amplitude that vanishes at
small
fixed angle like $\exp (-\theta\sqrt {\alpha' s}) $~\cite{oog}. Recently
this
approach was applied~\cite{okada} to the TeV string scenario.
The conclusions of ref.~\cite{grossmende} would seem to
disagree with
those of ref.~\cite{venezia}, based on the string eikonal, according to
which
the scattering at ${\sqrt s}\gg M_s$ is well described by (semi)classical
general relativity. However, as  explained in ref. \cite{oog},
the Borel resummation leading to the result $\exp (-\theta\sqrt {\alpha'
s}) $
applies without relevant corrections only in a limited range of energies
${\sqrt s}< M_s(\ln g_s^2)^{3/2}$. Indeed for energies such that
$\lambda_s
\gg R_S$ the two approaches give results that are fairly consistent with
each
other~\cite{veneziano}. So we believe that in the transplanckian
regime our simple approach is  valid. Notice, that also in the approach of
 ref.~\cite{venezia} the elastic amplitude has an exponential
suppression, different  than that in refs.~\cite{grossmende,oog}, and
 due to gravitational radiation and
diffractive string production. Moreover the discussion of
ref.~\cite{venezia}
shows that, even if the elastic channel is suppressed, the total cross
section at finite angle is large and grows with $\sqrt s$.
Notice also that if $\sqrt s$
is large enough the average mass of the diffractive states eventually
exceeds both
$M_S$ and $M_D$, so it is  possible that the diffractive channel
becomes at extremely high energies the production of a pair
of  small black holes.

Next, we would like to comment on the possible effects of the modes 
corresponding to the
brane motion, the branons~\cite{Sundrum}. 
The first comment is that these modes do 
not necessarily exist.
If the SM fields are localized at some fixed point, as in orbifold 
constructions, then it is consistent to assume
that the corresponding ``brane'' is not a dynamical object, and could also have
vanishing tension. So when the SM lives at a fixed point there is no effect 
(apart from a 
``trivial'' reduction of the phase space of bulk gravitons).

Two types of effects can be associated with brane dynamics. 
First, the presence
of a brane tension $\tau=\mu^4$ gives rise to a gravitational field, 
which at a tranverse distance
$y$ from the brane vanishes like $(R_\tau/y)^{n-2}$, where 
the gravitational radius $R_\tau$ is~\cite{empa} 
\beq
R_\tau^{n-2}=\frac{\Gamma(n/2) 2^{n+1} \pi ^{\frac{n}{2}}}
{{\sqrt{(n-1)(n+2)}}}\frac{\tau}{M_D^{2+n}}.
\eeq
The wave function of a bulk graviton with transverse momentum $k_T$ is 
expected to receive
a correction of order $ (R_\tau k_T)^{n-2}$  from the presence of 
this background. For a given impact
parameter $b$, the tree-level scattering amplitude
is precisely saturated by the exchange of graviton KK modes with mass $\sim
1/b$. Therefore we expect the background metric to correct the eikonal phase
by a relative amount $\sim (R_\tau /b)^{n-2}$. For $n>2$, 
at large enough impact parameter we
can neglect the brane radius as much as we can neglect the Schwarzschild radius $R_S$. 
This gravitational brane 
effect becomes less important as the tension is decreased. 

 \begin{figure}
 \centering
 \epsfysize=1.0in
 \hspace*{0in}
 \epsffile{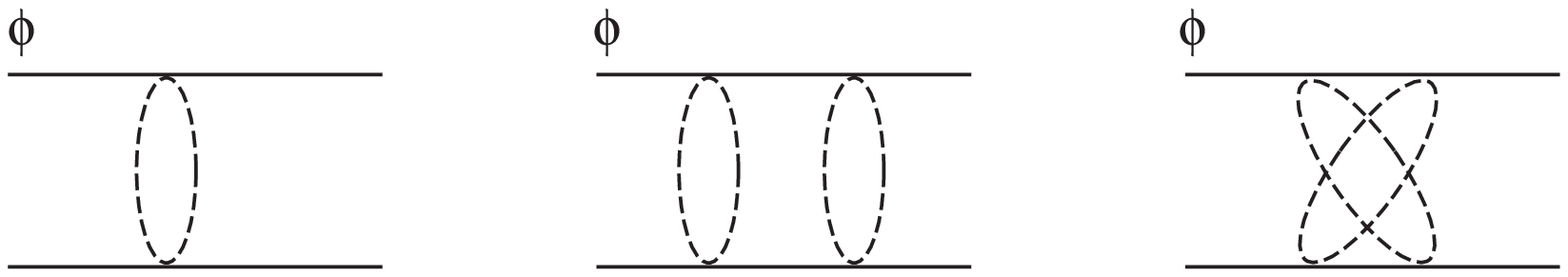}  
 \caption{Ladder and cross-ladder Feynman diagrams contributing
to elastic scattering of $\Phi$ particles. The dashed lines represent
the exchange of brane excitations.}
 \label{figb}
 \end{figure}

The second class of effects is associated with the exchange of branons. 
Branon exchange
becomes more important the smaller the tension. Notice first of all that at the
classical level branons are not excited by the gravitational 
shockwave of a fast
moving particle, {\it i.e.} the induced metric on the brane remains of the
Aichelburg-Sexl kind.
At the quantum level, branons can be exchanged in pairs between the colliding
particles. The exchange of a pair of branons is very much equivalent to the exchange 
of a``composite'' graviton. In order to get an estimate of the size of 
the effect, 
we may generalize the graviton eikonal including the ladder and cross-ladder
diagrams with
two-branon blobs (see fig.~\ref{figb}). 
The eikonal phase gets modified into
\beq
\chi(b)=\left(\frac{b_c}{b}\right)^n + \frac{n}{120\pi^3}\frac{s}{\tau^2 b^6}.
\label{braneik}
\eeq
In the case of $n=6$ the effect is equivalent to a 
renormalization of $G_N$, but in all
other cases it can distort our result from pure graviton exchange. 

Including both effects from brane excitations,
we can write the relative correction to the elastic amplitude (in the semiclassical region) as
\beq
\frac{\delta {\cal A}}{\cal A}= \left (\frac{\mu}{M_D}\right )^4 \left 
(\frac{M_D{\sqrt -t}}{s}
\right )^\frac{n-2}{n+1}+
\left (\frac{M_D}{\mu}\right )^8 \left (\frac{M_D{\sqrt -t}}{s}
\right )^\frac{6-n}{n+1},
\eeq
where we have dropped all numerical factors.
This equation shows that for the interesting case $2<n\leq 6$ and for the 
natural choice $\mu\sim M_D$, both branon effects are small in 
the region of small angle
and large $s$. Notice that for a 3-dimensional D-brane arising 
by wrapping a $(9-n)$--brane over
the $6-n$ dimensions of radius $\sqrt {\alpha'}$ along the lines of the example
above, the tension is given by
\beq
\tau=\frac{e^{-\phi}}{(2\pi)^{\frac{n}{2}} {\alpha'}^2}.
\eeq
Then at weak coupling we have $R_\tau<\sqrt {\alpha'}$ 
and the effects of the first class are
very small. Effects of the second class are also small as long as 
$b>\sqrt {\alpha'}$
and $n<6$. For $n=6$, the contribution from branon exchange leads to
the peculiar result $\chi_{\rm branon}=\chi_{\rm grav}/5$, see 
\eq{braneik}\footnote{We thank A. Strumia for 
stressing this result of the physics of D-branes.}.
 
To summarize, quantum-gravity corrections are potentially 
very sizable because, although
they involve smaller length scales, they do not appear with large exponents,
as in the case of the semiclassical corrections. Of course, they cannot
be computed in the absence of a complete quantum-gravity theory. We have
estimated their effects in the context of string theory, borrowing the
results of ref.~\cite{venezia} (which are however strictly valid only for 
graviton-graviton scattering). For $b<b_D$, a diffractive production of
string states occurs, depleting the elastic scattering cross section,
although it may still be possible to recover the gravitational scattering 
properties by studying inclusive cross sections. This gravitational
excitation of string modes looks like a very promising way to investigate,
in a rather clean channel, string effects at high-energy colliders. As the
impact parameter is further decreased and becomes of the order of
$\lambda_S=\sqrt{\alpha'}$, we are entering the regime of head-on collisions
between string modes, with inelastic production of Regge excitations
eventually leading to multi-string 
states~\cite{dimemp}.

In the following, we will focus on gravitational transplanckian collisions,
neglecting quantum-gravity effects. One should keep in mind
that such (at the moment incalculable) effects may 
give significant modifications to the signal discussed below.
Although potentially disrupting to the predictivity of the elastic channel,
these new-physics effects are of course very interesting from the
experimental point of view.

\section{Phenomenology in the Transplanckian Regime}
\subsection{Signals at the LHC}
\label{sec3}

At the LHC, the observable of interest is jet-jet production at small
angle (close to beam) with large center-of-mass collision energy.
The amplitudes derived in previous sections are applicable for the 
scattering of any two partons.  The total jet-jet 
cross-section is then obtained
by summing over all possible permutations of initial state quarks
and gluons, using
the appropriate parton distribution weights and enforcing kinematic
cuts applicable for the eikonal approximation.

Defining $\hat s$ and $\hat t$ as Mandelstam variables of the parton-parton
collision,  
we are interested
in  events that have $\sqrt{\hat s}/M_D\gg 1$ and 
$-\hat t/\hat s \ll 1$. We can extract $\sqrt{\hat s}$ 
from the jet-jet invariant mass $M_{jj}=\sqrt{\hat s}$, 
and $\hat t$ from the rapidity separation of
the two jets $-\hat t/\hat s=1/(1+e^{\Delta \eta})$, where 
$\Delta \eta \equiv \eta_1-\eta_2$.  The variable $\Delta \eta$ is
especially useful since it is invariant under boosts along the beam
direction, and it is simply related
to the $\hat \theta$ scattering angle in the center-of-mass frame:
\beq
\Delta \eta = \ln\left[ \frac{1+\cos\hat\theta}{1-\cos\hat\theta}\right].
\eeq
Therefore, the kinematical region of interest is defined by
the equivalent statements 
\beq
\Delta\eta\to \infty ~~~\leftrightarrow ~~~ 
\hat\theta \to 0 ~~~ \leftrightarrow ~~~
\frac{-\hat t}{\hat s}\to 0.
\eeq

In computing the cross-sections 
we use the CTEQ5~\cite{cteq} parton-distribution functions.
We have also compared results with GRV~\cite{grv} and find little difference.
We evaluate the parton-distribution functions at the scale $Q^2=b_s^{-2}$, see
\eq{defbs}, if $q>b_c^{-1}$
and $Q^2=q^2$ otherwise ($q^2\equiv -\hat t$) \cite{ratta}.

We require that both jets have $|\eta|<5$ 
and $p_T>100\gev$ from conservative detector requirements. 
The SM di-jet cross section is computed using Pythia~\cite{pythia},
ignoring higher-order QCD corrections, and the gravitational
signal is calculated using our own VEGAS Monte Carlo integrater.  
For simplicity
we are defining the background as the jet-jet cross-section from QCD with
gravity couplings turned off, and the signal as the jet-jet cross-section
from the eikonal gravity computation with QCD turned off.  
In reality, 
SM and gravity contributions would be
simultaneosly present. Nevertheless, there is no interference between
the leading QCD contribution in the limit $t/s\to 0$ and the gravitational
contribution, since graviton couplings are diagonal in color indices.
However, terms ${\cal O} (\alpha_W )$ are generated by the interference
between gravitational and $Z/\gamma$ contributions. At order 
${\cal O} (\alpha_s^2)$, the di-jet cross section receives contributions
from the square of the amplitude obtained by exchanging 1 gluon and any
number of gravitons (resummed in the eikonal approximation) and
from the interference between the gravitational amplitude and resummed diagrams
with exchange of 2 colored particles and any
number of gravitons.
In the small regions
of parameter space where the QCD and gravity contributions are comparable, 
these terms, which will not be computed in this paper, should
be taken into account.  We will not make any detailed claims about
that region here.

Although we are working in the transplanckian regime, we are considering
processes with low virtuality ($Q^2 \sim 1/b^2\ll M_D$) and,
as long as no new physics appear at momenta smaller than $Q$, the SM
dynamics give a reliable estimate of the background.
In the context of string theory, the QCD background has to be 
interpreted as the
leading contribution from exchange of open strings, in the limit in which
the transfer momentum is smaller than the mass of the Regge excitations. 

The jet-jet process is best described by the two kinematic 
variables  
$M_{jj}$ and $\Delta \eta$.  Although
experimental searches for new physics
in the jet-jet observable should investigate the full range of these two
kinematic variables, our goal is to study the region in which the 
eikonal computation is valid, where we can make reliable theoretical
predictions.
This leads us to the region of
large $\Delta \eta$ and large
$M_{jj}$.

Let us begin by studying  the parton differential cross section
signal distribution as
a function of the rapidity separation $\Delta\eta$ for a fixed
$\sqrt{\hat s}=M_{jj}$,
\beq
\frac{d\hat\sigma}{d\Delta \eta}=\frac{\pi b_c^4 \hat s e^{\Delta \eta}}{
\left( 1+ e^{\Delta \eta} \right)^2} \left| F_n (y)\right|^2,
\eeq
where $y=b_c \sqrt{\hat s}/\sqrt{1+e^{\Delta \eta}}$.
This distribution has a series of peaks whose maxima and minima
are determined by the values of $\Delta \eta$ that
satisfy the equation
\beq
1-e^{- \Delta \eta}=-\left. \frac{y}{|F_n (y)|}\frac{d|F_n (y)|}{dy}
\right|_{y=\frac{b_c \sqrt{s}}{\sqrt{1+e^{\Delta \eta}}}}.
\label{picco}
\eeq

These peaks arise from the oscillations of the function $|F_n|$ around
its asymptotic value given in \eq{statp}, and characterize the transition
region between the ``classical'' (small $\Delta \eta$) and
``quantum'' (large $\Delta \eta$) regimes discussed in sect.~\ref{sec2}.
Therefore, we cannot solve \eq{picco} using the asymptotic
expressions given in \eq{singg}
or \eq{statp}. Approximate solutions of \eq{picco} can be found if the
peaks are located at sufficiently large values of $\Delta \eta$, since
in this case
we can neglect $e^{- \Delta \eta}$ with respect to 1 in 
the left-hand side of \eq{picco}. Then, \eq{picco} becomes a function 
only of the variable $y$. For our consideration, this approximation
is adequate for determining at least the first peak. 
We find that the first peak of the $\Delta \eta$
distribution, for a fixed value of the two-jet invariant mass $M_{jj}$, 
is given by
\beq
\Delta \eta^{\rm (peak)} =\frac{2(n+2)}{n}
\ln  \left( \frac{k_nM_{jj}}{M_D} \right) .
\label{piccolin}
\eeq
The coefficients $k_n$ can be obtained by numerically solving
\eq{picco} and are given by 
$k_{2,3,4,5,6}=0.8,\, 0.9,\, 1.0,\, 1.2,\, 1.3$.

These peaks are a characteristic feature of the higher-dimensional
gravitational force, and correspond to the diffraction pattern of
the scattered particles. As discussed in sect.~\ref{physical},
higher-dimensional gravity, albeit leading to
a force with infinite range, defines a length scale ($b_c$) in the
quantum theory. A diffraction pattern emerges when 
$q\sim b_c^{-1}$ corresponding to the interference of the scattered
waves in the region
$b\sim b_c$.
In the Coulomb case
($n=0$), such a scale does not exist 
and therefore no diffractive pattern is produced.

The di-jet differential cross section $d\sigma_{jj}/d|\Delta\eta |$
is plotted in
fig.~\ref{fig-dely} for $n=6$, 
$M_{jj}>9\tev$ and $M_D=1.5\tev$ and $3\tev$.  Since
the parton-distribution functions 
decrease rapidly at higher $M_{jj}$, the value of
$M_{jj}$ in \eq{picco} is well approximated by $9\tev$ in this
example.  The first peaks are then calculated 
from \eq{piccolin} to be at $\Delta \eta =5.5$ ($M_D=1.5\tev$) and
$3.7$ ($M_D=3\tev$).  The Monte Carlo
integrated distributions have peaks that agree well with
these numbers.  

\jfig{fig-dely}{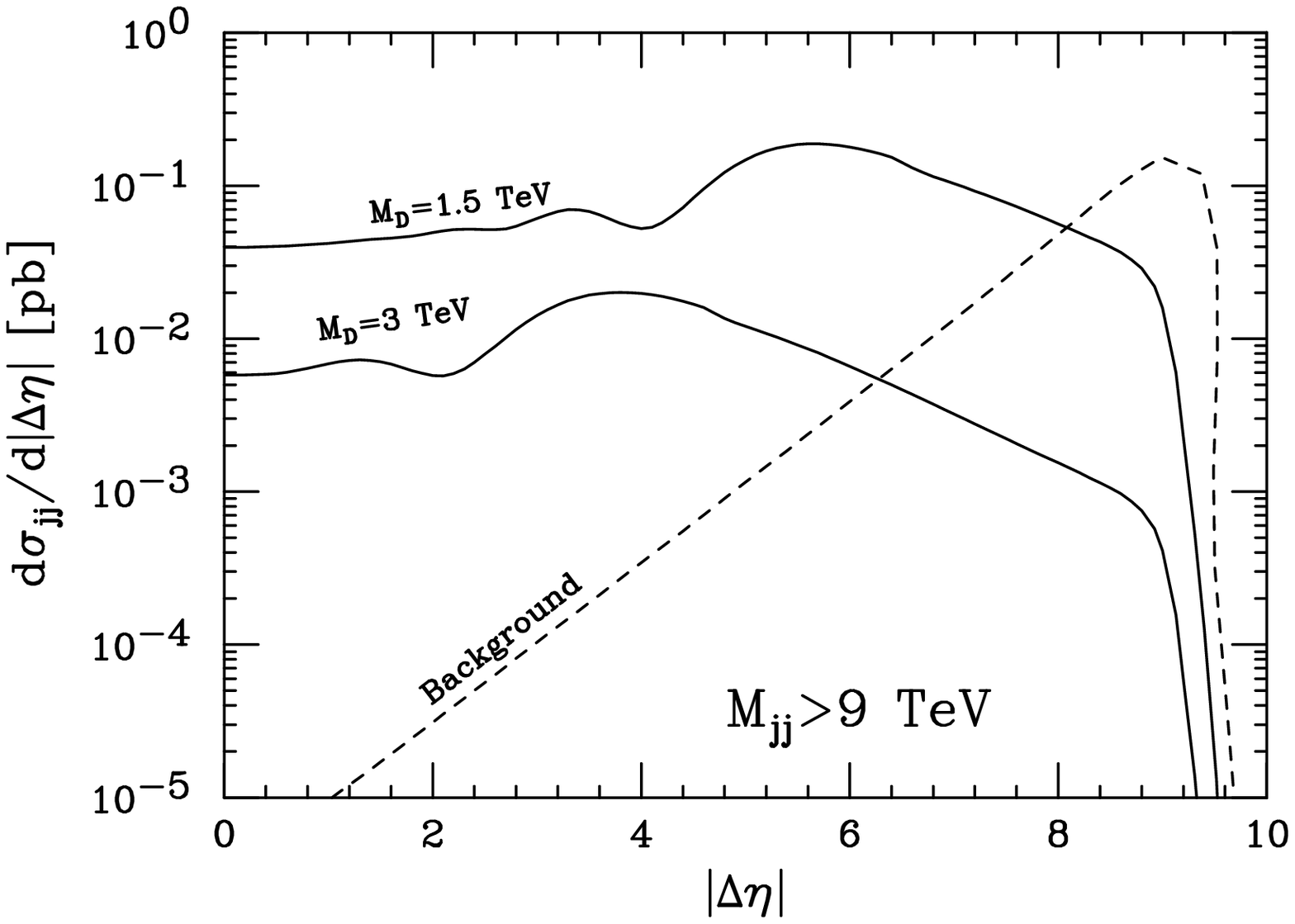}{The 
di-jet differential cross section 
$d\sigma_{jj}/d|\Delta\eta |$ from
eikonal gravity for $n=6$, 
$M_{jj}>9\tev$, when both jets have $|\eta | <5$ and $p_T>100\gev$,
 and for $M_D=1.5\tev$ and $3\tev$.  The
dashed line is the expected rate from QCD.}

The same differential cross section is shown in fig.~\ref{cycle-dely-n}
for different values of $n$ and for $M_D=1.5\tev$. For $n=2$ the
first peak is partly hidden by the logarithmic divergence for $\hat t
\to 0$, and no structure after the first peak is visible. As $n$
increases, the peaks become more evident. The
study of the peak structure could be a feasible experimental technique to
measure the number of extra spatial dimensions $n$. However, such a
study can only be pursued with a full detector simulation, taking into
account the rapidity and jet mass resolutions.

\jfig{cycle-dely-n}{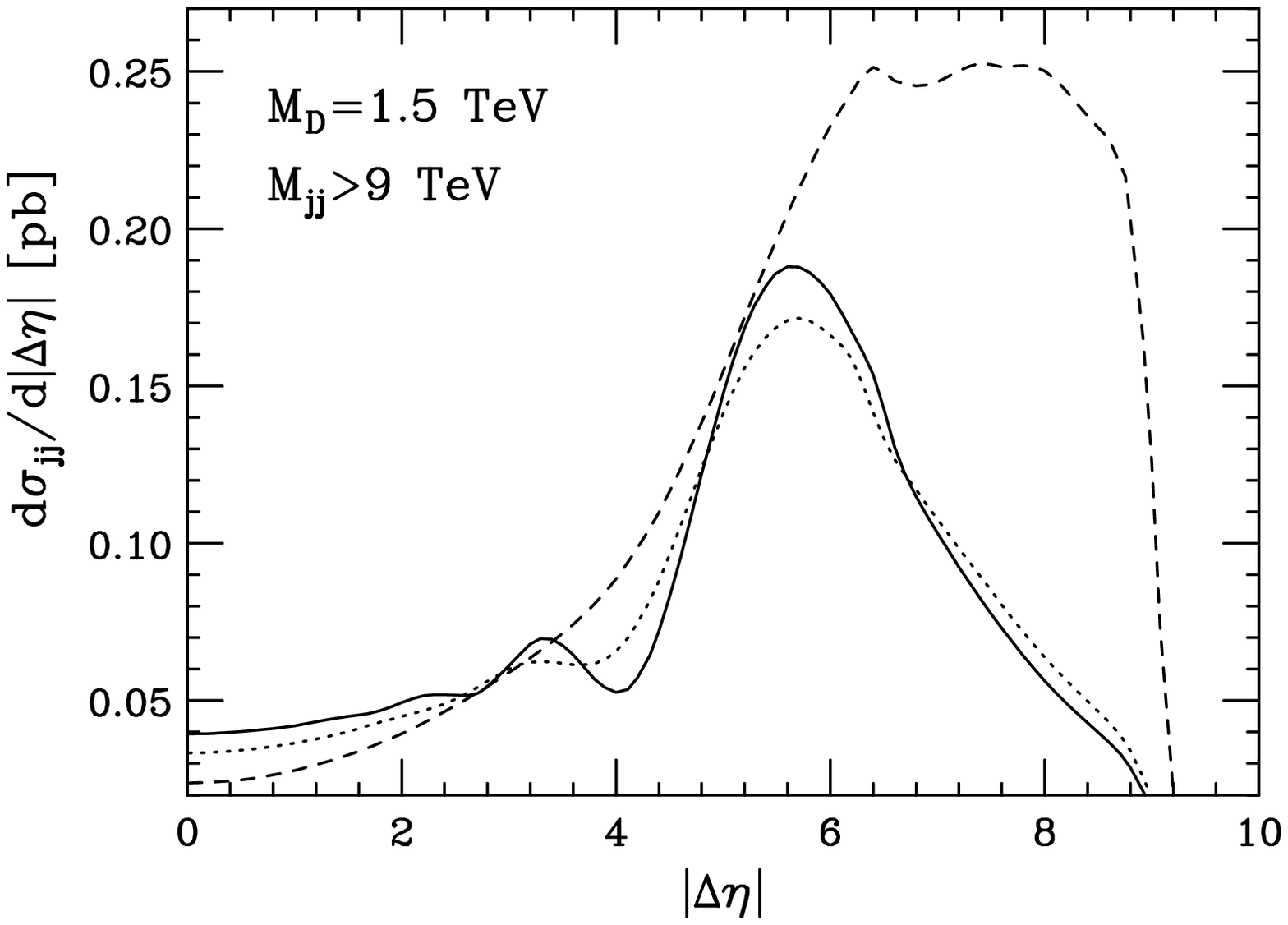}{The 
di-jet differential cross section 
$d\sigma_{jj}/d|\Delta\eta |$ from
eikonal gravity for $M_D=1.5\tev$ and $n=6$ (solid line), $n=4$ (dotted line),
and $n=2$ (dashed line). We require
$M_{jj}>9\tev$ and that both jets have $|\eta | <5$ and $p_T>100\gev$.}
 
Since the two jets are experimentally indistinguishable, we have
used $|\Delta \eta |$, instead of $\Delta \eta$, as the appropriate
kinematical variable to plot. This means that the experimental signal
considered here contains also contributions from scattering with large
and negative $\Delta \eta$, which corresponds to partons colliding
with large momentum transfer and retracing their path backwards.
For the background, these effects are calculable and taken into account.
However, the theoretical estimate of the signal at negative $\Delta\eta$
lies outside
the range of validity of the eikonal approximation. 
Nevertheless, this is expected to be negligible and
can be safely ignored. Indeed,
by dimensional arguments, the differential cross section for $t\to -s$
is estimated to be $d\sigma /dt \sim \pi R_S^2/s$, see \eq{diff2}.
When compared to the signal at small $t$ ($d\sigma /dt \sim \pi b_c^4$),
this gives a correction ${\cal O} [R_S^2/(sb_c^4)] \sim
{\cal O} [(M_D^2/s)^{\frac{(n+2)^2}{n(n+1)}}]$, which is smaller
than the terms we have neglected in our calculation, see \eq{correx}.
Notice that this complication is unavoidable in the presence of
identical colliding particles, since the amplitude is symmetric in
$t$ and $u$ ($u=-s-t$). 

Recalling our discussion of the corrections to the eikonal amplitude given
in sect.~\ref{seccoreik}, we notice that semiclassical corrections are indeed
small. In the kinematical region of interest, the second term in
\eq{correx} amounts to less than 1\%
for $M_{jj}=6M_D$ and $n<6$
and less than 5\% for $M_{jj}=3M_D$ and $n<6$, while the first term 
amounts to about 5\% at $\Delta\eta =3$. 
Gravitational radiation at $\Delta\eta =3$ and $M_{jj}=3M_D$
gives corrections of about 7\% for $n=2$ and 5\% for $n=6$, see \eq{imclass}.
Quantum gravity can however severely affect our signal. 
If the corresponding corrections had the form $(\lambda_P/b_c)^2$, then
they would amount to 5\% (6\%) for $M_{jj}=6M_D$ and $n=2$ ($n=6$) or
to 20\% (9\%) for $M_{jj}=3M_D$ and $n=2$ ($n=6$), and they would be 
under control. However, it is rather likely that the dynamics taming
the ultraviolet behavior of gravity sets in at energies lower than $M_D$,
enhancing these corrections. This is indeed the case of weakly-coupled strings,
where in the previous estimate $\lambda_P$ should be replaced by $\lambda_S$,
giving an enhancement of $M_D^2/M_S^2$.
For instance,
if the diffractive
string production discussed in sect.~\ref{seccor}
takes place, it will deplete the elastic channel at impact
parameters smaller than $b_D$, which is typically very close to $b_c$,
for values of $M_D$ relevant for the LHC and for weakly-coupled strings. 
In this case, string production
could be the discovery process at the LHC.
 
As is apparent from fig.~\ref{fig-dely},
the background increases with $\Delta \eta$ faster than the
signal, and so the best signal to background  ratio is found at the
smallest $\Delta\eta$.  However, small $\Delta\eta $ are not
in accord with the eikonal approximation.  To stay within the acceptable
kinematic range of the eikonal approximation
we impose the condition $\Delta\eta >3$, which is equivalent to
$-\hat t/\hat s <1/21$ and $\hat \theta <25^\circ$.
We then select a maximum value of $\Delta\eta $ in order to reject the
background. In our study we make a simple universal choice 
$\Delta\eta <4$, but in practice one could choose the maximum
$\Delta\eta $ using only the criterion of retaining enough events to be
detected.
We want to stress that this universal range of $\Delta \eta$ is chosen
just for illustration, since its optimal choice depends on $M_D$.
In particular, if $M_D$ is below about 2~TeV, visible signals 
certainly can be obtained with a more stringent lower limit on $\Delta \eta$,
therefore investigating a region of smaller scattering angles, where
the eikonal approximation is even more accurate.

Furthermore, with the current computation technology of QCD, the dijet rate 
in our kinematic configuration is not a precision observable to compare with
theory. This is especially true at large rapidity separation. However, our
gravity signal can swamp the QCD expectation most notably at smaller
rapidity separation, where the hard-scattering QCD rate is smaller and
more reliably known.  That is partly the reason why we choose for some
plots to illustrate the gravity signal in the modest 
interval $3< |\Delta\eta | <4$.

At very large rapidity separation, the QCD dijet observable may be
dominated by BFKL dynamics~\cite{bfkl}, 
in which case a decorrelation of the azimuthal
angles of the two jets ensues~\cite{decorrelation}. 
Such a decorrelation is not expected in the color-singlet
gravity signal. It may even
be possible to use this to distinguish QCD from gravity.
Similarly, the gravity signal is not expected to fill the central
region with soft gluons, and the lack of this central jet activity could be
an additional discriminating tool.

\jfig{fig-Mjj}{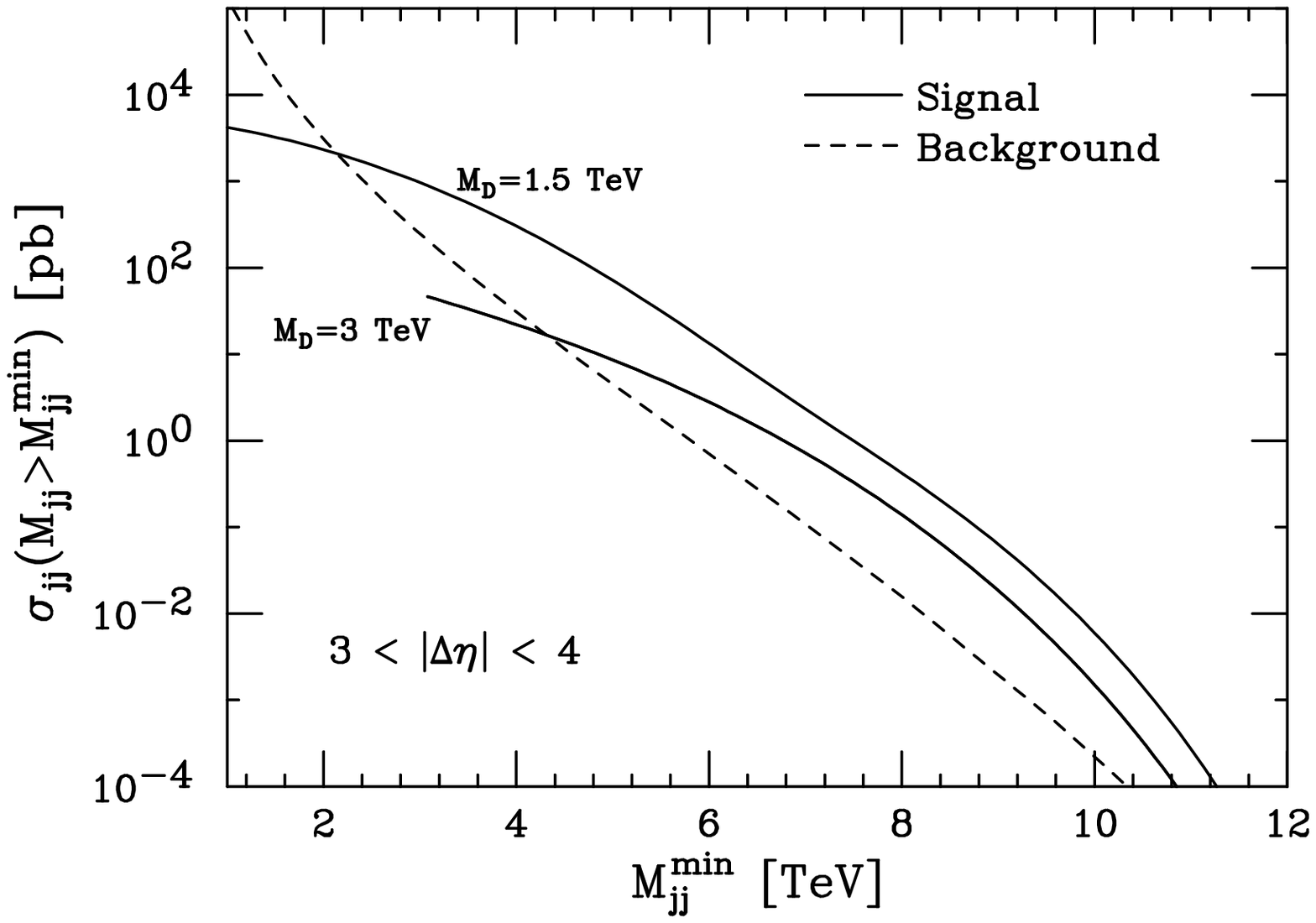}{Total 
integrated di-jet cross-section
for $3< |\Delta \eta| <4$, $n=6$, and $M_{jj}>M_{jj}^{\rm min}$,
when both jets have $|\eta | <5$ and $p_T>100\gev$. Lines are plotted
for $M_D=1.5$ and 3 TeV.
The eikonal approximation 
is reliable only where $M_{jj}/M_D\gg 1$. The expected QCD
rate is given by the dashed line.}

After having selected the range of $3< |\Delta \eta| <4$, we now
show in fig.~\ref{fig-Mjj} 
the cross-section as a function of minimum jet-jet invariant
mass cut for $M_D=1.5\tev$ and $3\tev$.  We plot results
for all $M_{jj}\geq M_D$, but we recall that the eikonal approximation is
valid only for $M_{jj}/M_D\gg 1$.  
This plot shows the important feature that
the signal cross-section is flatter in $M_{jj}$ than the background.
This enables better signal to background for larger $M_{jj}$ cuts,
which is the preferred direction to go for eikonal approximation
validity.  Therefore, one should make 
the largest possible $M_{jj}$ cut that still
has a countable signal rate for a given luminosity.

\begin{figure}[tpb]
\dofigs{3.3in}{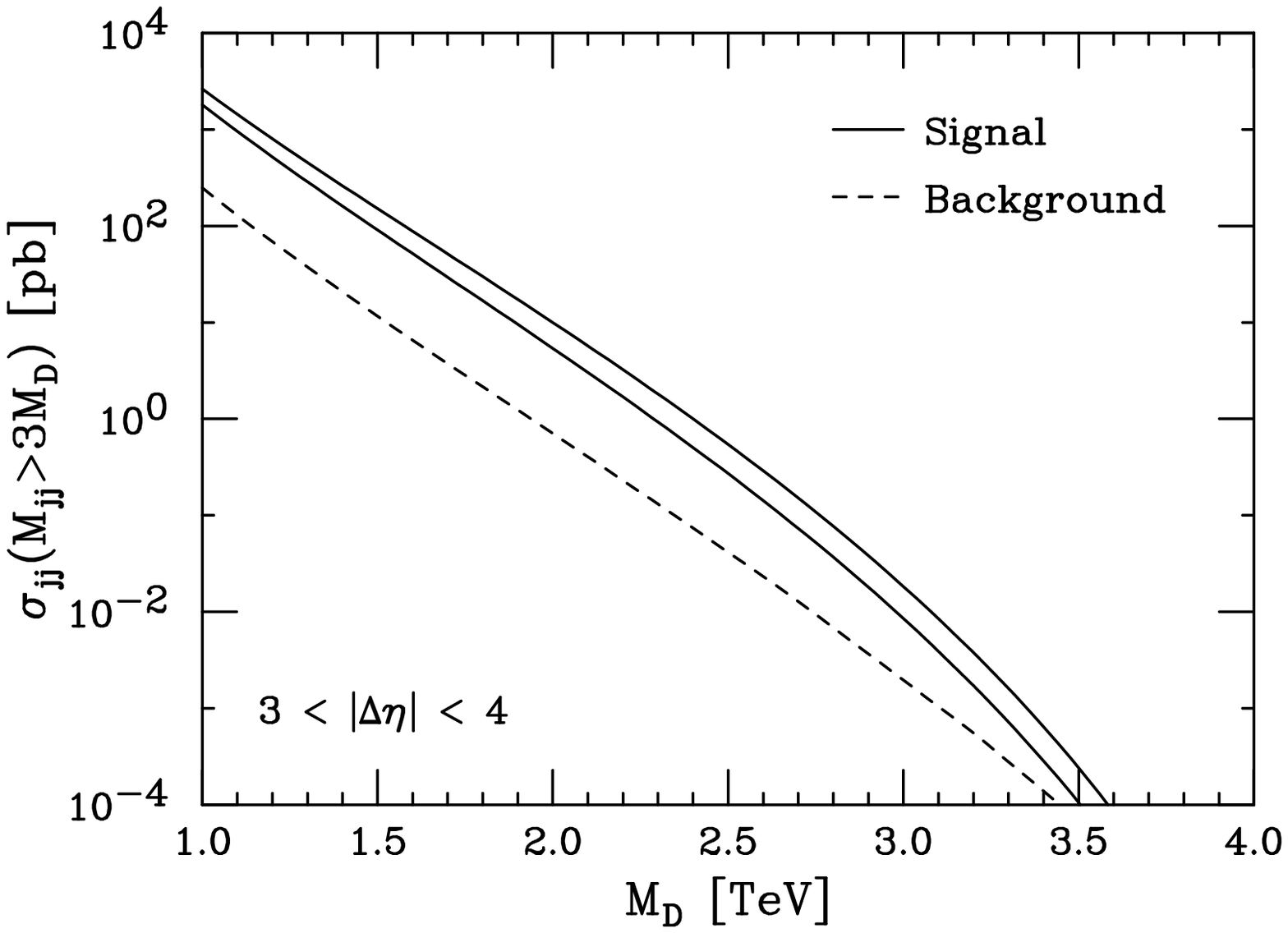}{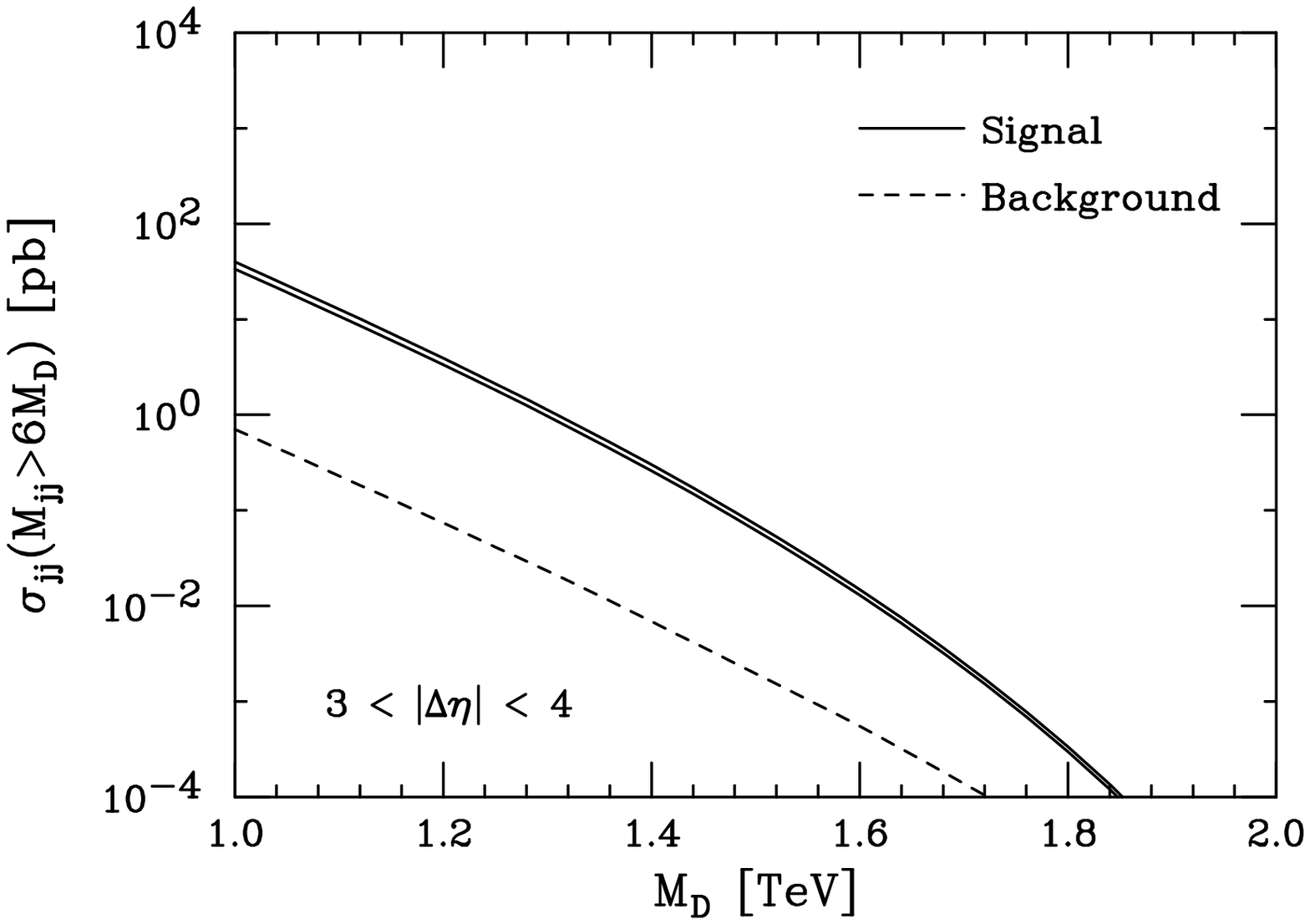} 
\caption{Total integrated di-jet cross-section for $3< |\Delta\eta | <4$,
when both jets have $|\eta | <5$ and $p_T>100\gev$,
for (left panel) $M_{jj}>3M_D$ and (right panel) $M_{jj}>6M_D$. In each
plot, the upper solid line is for $n=6$ and the lower solid
line is for $n=2$.  The dashed line is the expected QCD
rate.}
\label{fig-MD}
\end{figure}

Finally, in fig.~\ref{fig-MD} we plot the total integrated
cross-section as a function of $M_D$ for $M_{jj}>3M_D$ (left panel)
and $M_{jj}>6M_D$ (right panel).  
We also have required $3<|\Delta \eta |<4$.
The two solid lines correspond to $n=6$ (upper line) and $n=2$ (lower
line).  They are not far separated in this log plot.
Figure~\ref{fig-MD} demonstrates several universal features. First, larger
$M_{jj}/M_D$ corresponds to a larger ratio of gravity signal
to expected QCD rate, and to a more reliable applicability of
the eikonal approximation.  Larger $M_{jj}/M_D$ also means less range of
$M_D$ probed at the LHC since parton luminosity drops rapidly as
$M_{jj}$ approaches the $14\tev$ collider limit.
In this paper, we will restrict our considerations to
$M_{jj}/M_D>3$ and
so elastic scattering with $M_D$ as large as $3.5\tev$ can be tested 
with more than 10 events
in $100\xfb^{-1}$.  If we adopted the more conservative constraint
$M_{jj}/M_D>6$, the range of $M_D$ where large and calculable signals 
are expected reduces
to $1.8\tev$.

Before concluding this section on the LHC signatures, we recall
that the eikonal jet-jet observable analyzed here is not the only
transplanckian
non-SM process. Black-hole production is also expected to be very large
with a parton-parton production cross-section given approximately
by $\sigma_{BH} \simeq \pi R_S^2$.  As discussed earlier, this is
just a dimensional-analysis estimate
and reflects only the expectation that all collisions
with impact parameter $b<R_S$ get absorbed into a black hole.  

\begin{figure}[tpb]
\dofigs{3.3in}{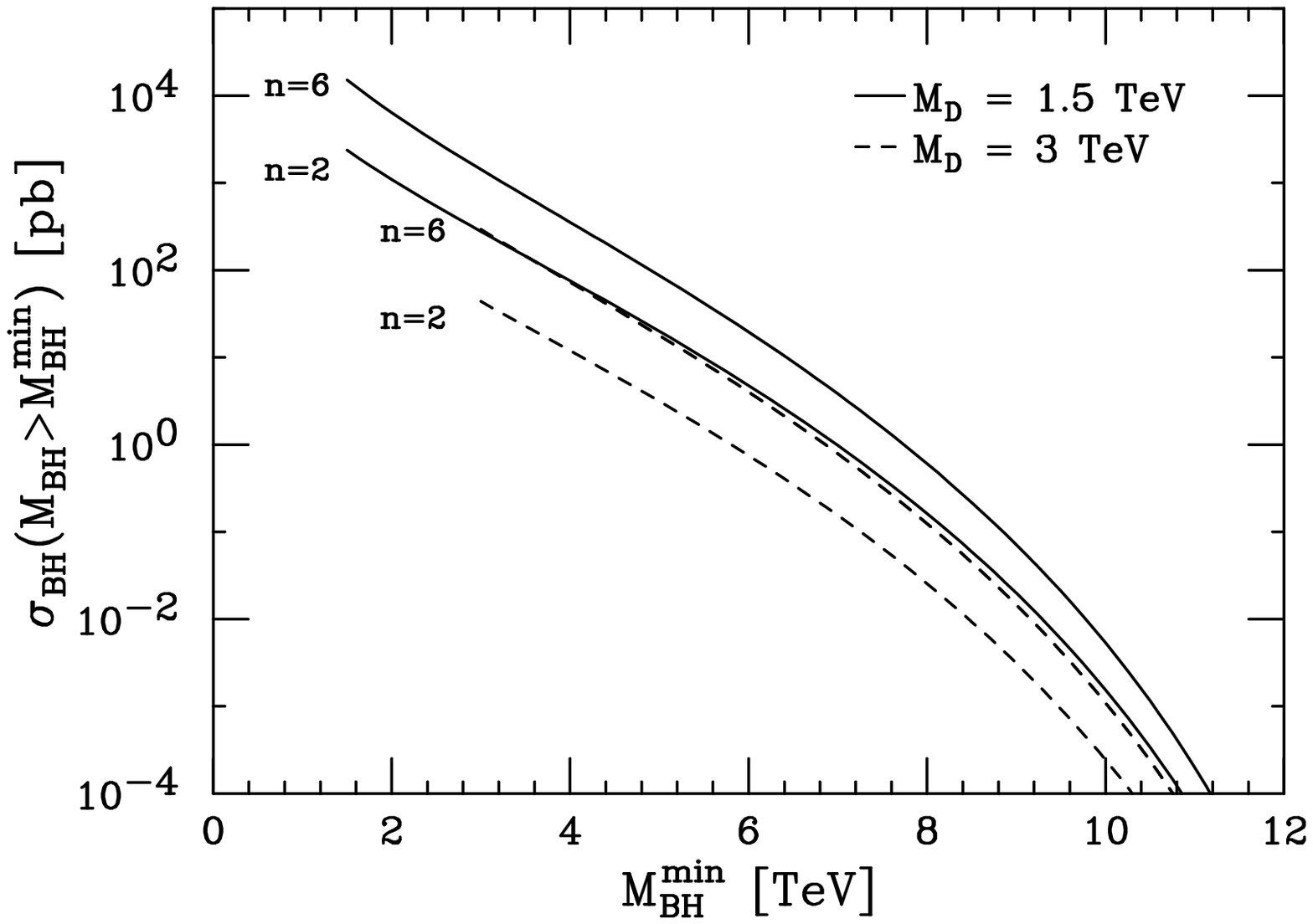}{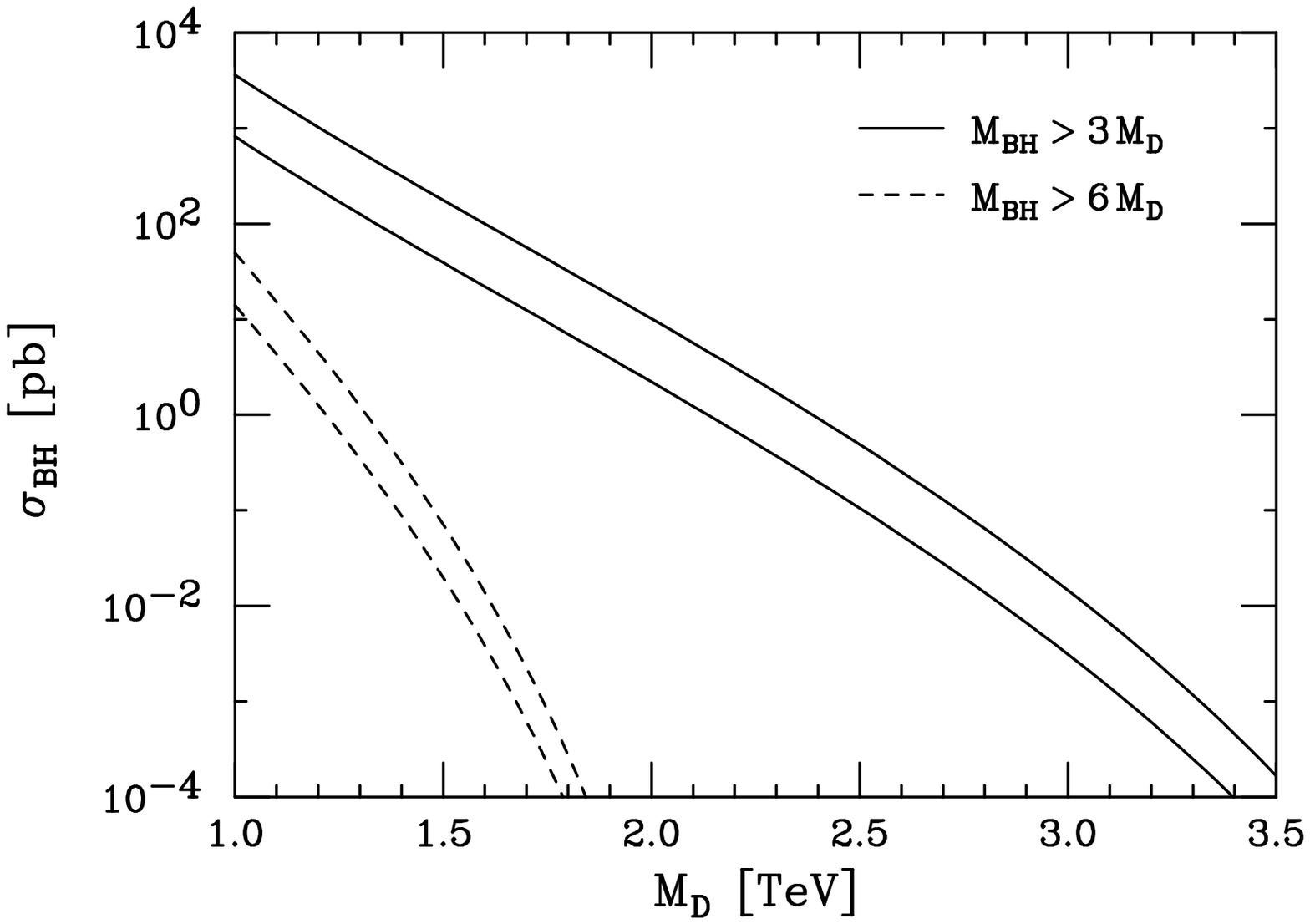} 
\caption{Black-hole production cross-section as a function of
(left panel) minimum black-hole mass and (right panel) the gravity
scale $M_D$.}
\label{fig-BH}
\end{figure}

The description of the scattering process in terms of 
black-hole production is expected
to be reliable only for black hole masses $M_{\rm BH}$
well above the $D$-dimensional gravity scale $M_D$.  In the
left panel of fig.~\ref{fig-BH}
we plot the production cross-section as a function of minimum black-hole
mass for $M_D=1.5\tev$ and $M_D=3\tev$.  
Although the lines are extended down to
$M^{\rm min}_{\rm BH}= M_D$,  
the reliable region is only $M_{\rm BH}/M_D\gg 1$.  The signatures
of black holes are spectacular high-multiplicity events with almost
no SM background~\cite{blackh}.
Only a few events at very high invariant mass are needed to
identify a signature.  The right panel of fig.~\ref{fig-BH} shows the
cross-section for black-hole production for $M_{\rm BH}>3M_D$ and 
$M_{\rm BH}>6M_D$.  
With an integrated luminosity of $100\xfb^{-1}$,
a minimum of several events can be achieved
for $M_D\simeq 3.5\tev$, if we require $M_{\rm BH}>3M_D$, or for
$M_D\simeq 1.8\tev$ if $M_{\rm BH}>6M_D$.
These values of $M_D$ are nearly identical to what one can reach with
the jet-jet observable discussed above.  The combined search
for black-hole production and small angle jet-jet events will
be important for a full experimental characterization of the 
extra dimensions and for our ability to determine the underlying parameters
of the theory.

\subsection{Other Future Colliders}

In addition to Tevatron upgrades and LHC, there
are many colliders envisioned for the future.  These include
variants on the $e^+e^-$ linear collider theme,
the muon collider, and the very large hadron collider (VLHC).  Observables
associated with gravity are very sensitive to collider energies.  The search
for phenomena of almost all beyond-the-SM schemes is interested in higher
and higher energies, but gravity is especially dependent on this progress.
The reason is that either 
the rate increases at a high power of the collision energy if $s\ll M_D^2$,
(\eg,  $\sigma \sim s^{n/2}/M_D^{n+2}$ for jet plus graviton signatures); or,
well-chosen observables become calculable if $s\gg M_D^2$, (\eg, in the
case of eikonal gravity jet-jet production). Such 
calculability is taken for granted
in many other beyond-the-SM frameworks.
Either way, the highest energies attainable are important for increased
understanding if low-scale gravity is correct.

We would therefore like to briefly discuss transplanckian collisions in
the context of two of the highest energy colliders
being considered in the not-too-distant future.  The first is the 
VLHC~\cite{vlhc},
whose center-of-mass energy for proton-proton collisions is envisaged
to be between 50 TeV and 200 TeV. Obviously the higher energy of
this post-LHC machine will be much more probing. Much of the analyses that
we have described for the LHC would be applicable for VLHC analyses.
By going to higher center-of-mass energies, we can gain in the range
of sensitivity to $M_D$ and, more importantly, we can study kinematical
regions where quantum-gravity effects are expected to be smaller
and the eikonal approximation becomes more trustworthy.
For instance, by analysing dijet final states at the VLHC with invariant
masses larger than 150~TeV, we can make quantum-gravity corrections
(proportional to $\lambda_S^2/b_c^2$), 
for a fixed value of the string scale, about a factor of 200 (for $n=2$)
or 6 (for $n=6$) smaller than their expected size in LHC experiments. 

The other collider discussed at the highest currently imagined energy is
CLIC~\cite{clic}, which is a 
two-beam $e^+e^-$ linear collider design.  Energies
as high as $\sqrt{s}=10\tev$ are being considered.  Although such
high-energy machines will have some spread in luminosity as a function
of the center-of-mass energy, one still expects that approximately
$1/4$ of the beam luminosity will be within 1\% of designed center-of-mass
energy.  With large gravity-induced cross sections ($\sim$pb or more) 
for all $M_D/\sqrt{s}\ll 1$, combined with relatively low background
($\sim$fb)
and impressive luminosity design goals of $1\, {\rm ab}^{-1}$,  CLIC
could certainly improve, verify and even discover the gravitational
origin of beyond the SM signatures.

The maximum values of $M_D$ that can be studied at CLIC in the 
gravitational deflection process (Bhabha scattering) are uniquely
determined by the condition for validity of the eikonal approximation.
For comparison, the condition equivalent to the one chosen in our
LHC analysis is $M_D< \sqrt{s}/3$. Therefore, a linear collider with
$\sqrt{s}=10$~TeV could probe a parameter-space region, where theoretical
calculations are reliable, which is very similar to the one that can
be studied by LHC. However, the cleaner $e^+e^-$ environment offer
several advantages for precision tests and parameter determinations.

In fig.~\ref{clic-theta} we give one example of the signal
and background distributions in the scattering angle $\theta$ for $M_D
=2$~TeV and
$\sqrt{s}=10\tev$.
Here $\theta$ is the scattering  angle of the electron
in the Bhabha process, and 
$\theta=0$ indicates the electron going down the beam pipe undeflected
by the collision.  
Notice that the SM background is completely insignificant as long as
we exclude a small region around $\theta =0$. This allows experimental
studies of the cross section in a much more forward region than what is
possible at the LHC.

\jfig{clic-theta}{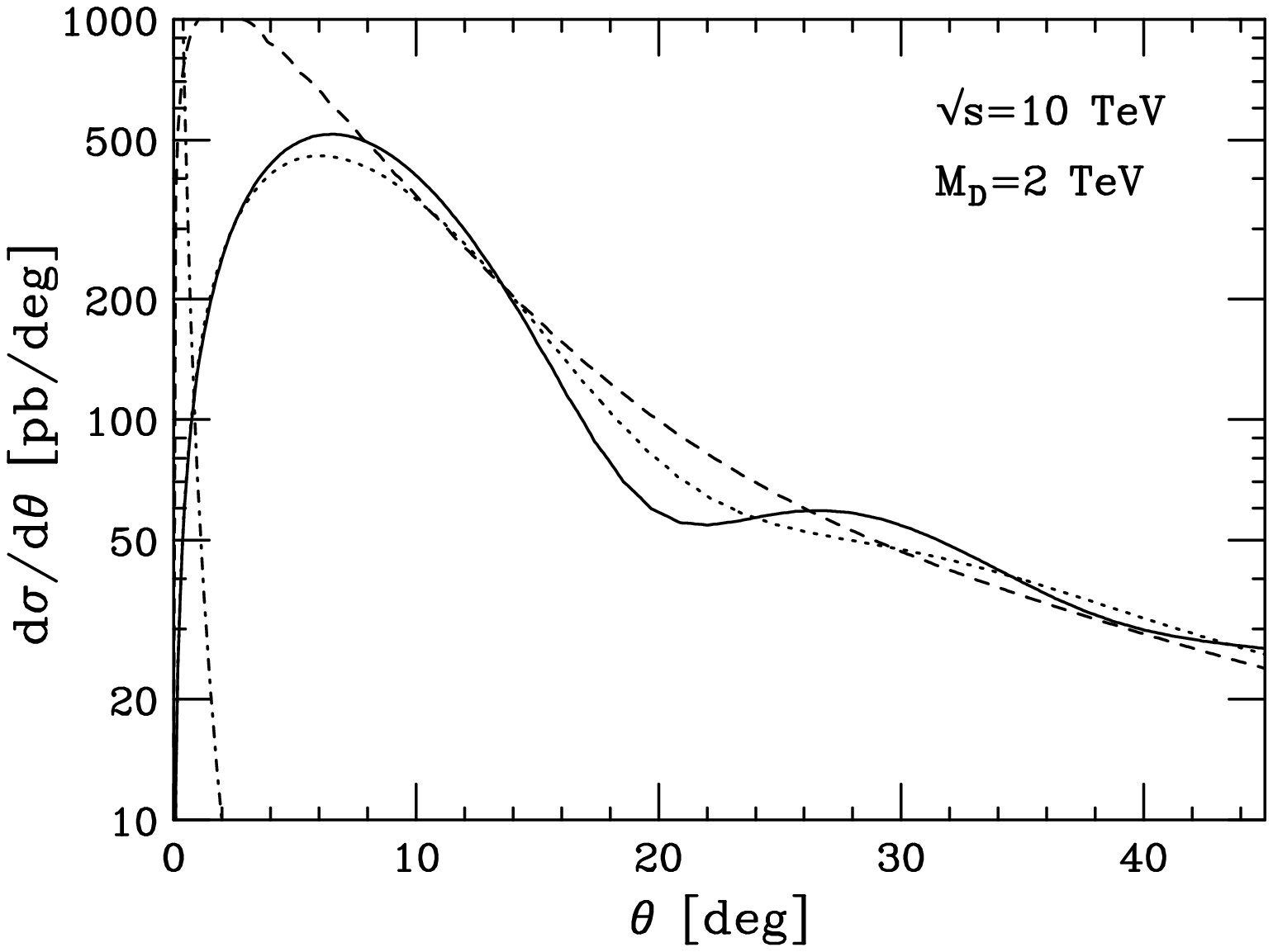}{ 
Angular distribution of the
$e^+e^-\to e^+e^-$ signal for $M_D=2\tev$
and $\sqrt{s}=10\tev$.  The solid line is
for $n=6$, the dotted for $n=4$ and the dashed for $n=2$.  The almost-vertical
dash-dotted line at the far left of the figure
is the expected Bhabha scattering rate from the Standard
Model.}

The signal angular distribution is characterized by the peak structure
encountered before, with minima and maxima described by the equation
\beq
\frac{1}{2} \left(1 -\tan^2 \frac{\theta}{2} \right)
=-\left. \frac{y}{|F_n (y)|}\frac{d|F_n (y)|}{dy}
\right|_{y=b_c \sqrt{s}\sin \frac{\theta}{2}}.
\eeq
The first peak  is approximatly given by
\beq
\theta^{\rm (peak)} \simeq 
\left( \frac{a_nM_D}{\sqrt{s}}\right)^{\frac{n+2}{n}},
\eeq
where $a_n$ is a numerical coefficient with value 
$a_{2,3,4,5,6}=0.9,\, 1.2,\, 1.1,\, 1.1,\, 1.0$.
The $n=6$ line (solid line) in fig.~\ref{clic-theta} is again the most telling
line, since a careful measurement of the ups and downs of
$d\sigma/d\theta$ would be hard to reproduce in another framework.
Given the good energy and angular resolutions that can be achieved at a
linear collider, CLIC could perform a much more precise study of the
peak structure than what is feasible at the LHC.

Another advantage of CLIC is that unlike a $pp$ collider, the 
scattering particles are distinguishable.  This means that the
calculable limit of $t\to 0$ (small angle scattering) is
unambiguously identifiable.  Recall that in the $pp$ collider case,
the case of partons glancing off each other at $\hat t\to 0$ was
not distinguishable from partons bouncing backwards at large
momentum transfer $\hat t\to -\hat s$.  We argued
that these non-calculable large momentum transfer contributions
are negligible, but CLIC can test that assertion.

\subsection{Comparison among Gravity Signals at the LHC}

Let us conclude by comparing the various signals and experimental 
strategies for
the discovery and the study of higher-dimensional gravitational interactions
at the LHC. The two relevant parameters are the fundamental Planck mass
$M_D$ and the number of extra spatial dimensions $n$. We are tacitly
assuming that the new dynamics of quantum gravity does not introduce a new 
mass scale smaller than $M_D$, or new phenomena in the relevant 
kinematical region.
The searches for
gravity signals at the LHC are classified in terms of three 
different kinematical regions.

{\it Cisplanckian Region.} This is the region in which the center-of-mass
energy of the parton collision is smaller than the Planck mass, $\sqrt{\hat s}
\ll M_D$. The theory can be described by an effective field-theory
Lagrangian with
non-renormalizable interactions, and the seach for new contact interactions
is an adequate experimental tool. However, the relations between the
coefficients of the contact interactions  and the fundamental gravity
parameters is model dependent. In this region, the search for jet plus missing
energy events is particularly interesting, because graviton emission can
be reliably calculated with a perturbative expansion. The region~\cite{noi} 
in which 
we can trust the effective theory for graviton emission
is $M_D>3.8$~TeV (for $n=2$) or $M_D>4.8$~TeV (for $n=4$). LHC is sensitive
to the graviton signal up to $M_D<8.5$~TeV (for $n=2$ and an integrated
luminosity ${\cal L}=100$~fb$^{-1}$) or $M_D<5.8$~TeV (for $n=4$ and 
${\cal L}=100$~fb$^{-1}$). In this window, LHC can perform a quantitative
test of high-dimensional gravity in the cisplanckian region.

{\it Planckian Region.} This is the intermediate region $\sqrt{\hat s}
\simeq M_D$, where no experimental signals can be predicted without
knowledge of the quantum-gravity theory. Signals from graviton emissions
or elastic gravitational scattering are also
present, but their rates cannot be reliably computed.  New and 
unexpected phenomena will be the key to understand the underlying
dynamics. Of course this is the region which will  eventually yield the
crucial experimental information.

{\it Transplanckian Region.} This is the region discussed in this paper,
characterized by $\sqrt{s}
\gg M_D$. The two experimental signals of interest 
in the transplanckian region 
are di-jet events (from elastic
parton scattering) and black-hole production. 
Unfortunately, in the case of the LHC we are rather limited 
in energy for both processes. The actual conditions for 
calculability we have taken are
$\sqrt{\hat s}
>6 M_D$ (in which the transplanckian region extends up to about 1.8~TeV)
or the more optimistic $\sqrt{\hat s}
>3 M_D$ (in which the transplanckian region extends up to about 3.5~TeV).
In both cases we are only marginally inside the transplanckian region 
since a complete separation between the quantum-gravity scale
$\lambda_P$ and the classical \sch radius $R_S$ has not been fully reached.
Therefore quantum-gravity contributions can potentially modify in a 
significant way both elastic
scattering and black-hole formation. In our analysis, we have assumed
that such contributions are small. 

While the black-hole
production cross section can only be estimated by dimensional analysis,
the elastic cross section in the small-angle region can be computed
as a perturbative expansion over controllable parameters. 
The elastic
cross section is larger than the one for black-hole production. 
The observation of a cross section
at finite angle growing with a power of $s$ would be a clean signal that 
the high-energy
dynamics of gravity has been detected.
Given
the highly characteristic events from black-hole evaporation, we expect
negligible backgrounds for the black-hole events. As we have shown in this
paper, the QCD background for di-jet events can be overcome by
studying the distributions in di-jet invariant mass and rapidity separation.
Therefore, the di-jet signal from
the gravitational deflection of partons is not limited by the background.
Its characteristic distributions can be used to test the gravitational
nature of the interaction and to determine the parameters of
the underlying theory.

{\bf Acknowledgments} We wish to thank I.~Antoniadis, J.~Barbon, J.~Bartels, 
S.~Catani, P.~Creminelli,
A.~De~Roeck, R.~Emparan, A.~Ringwald, A.~Strumia and G.~Veneziano 
for useful discussions.

\section*{Appendix}

In this appendix we collect some mathematical formul{\ae} which are useful
to reproduce computations presented in this paper.
The Bessel functions $J_n(x)$ have the following series expansions
\beq
J_n(x)=\sum_{k=0}^\infty \frac{(-)^k (x/2)^{2k+n}}{k! \Gamma (k+n+1)},
\eeq
\beq
J_n(x)=\sqrt{\frac{2}{\pi x}}\left\{ \cos \left[ x-(2n+1)\frac{\pi}{4}\right]
-\frac{4n^2-1}{8x}\sin \left[ x-(2n+1)\frac{\pi}{4}\right]
+{\cal O} \left( \frac{1}{x^2}\right) \right\}.
\eeq
Some useful integrals are
\beq
\int d^ny~e^{i{\vec x}\cdot{\vec y}} f(y)=
\frac{(2\pi)^{\frac{n}{2}}}{x^{\frac{n}{2}-1}}\int_0^\infty
dy~ y^{\frac{n}{2}}J_{\frac{n}{2}-1}(xy)f(y),
\eeq
for a generic function $f$, and with $x\equiv |\vec{x}|$, $y\equiv |\vec{y}|$.
\beq
\int_0^\infty dx~ x^a J_n (x) =2^a \Gamma \left( \frac{1+n+a}{2}\right)
/ \Gamma \left( \frac{1+n-a}{2}\right) .
\eeq
\beq
\int_0^\infty dx ~x^{a-1} \left( e^{ix^{-n}}-1\right) =
-\frac{1}{a}\Gamma \left( 1-\frac{a}{n}\right) e^{-\frac{i\pi a}{2n}}.
\eeq
\beq
 \int_0^\epsilon dx~x^a ~e^{ix^{-n}} =
\frac{\epsilon^{a+1+n}}{n} e^{i( \epsilon^{-n} +\frac{\pi}{2})}
\left[ 1+{\cal O}\left( \epsilon^n\right)\right]
~~~~~{\rm for}~~a>-n-1.
\eeq


\end{document}